\documentclass[12pt]{article}
\usepackage{xcolor}
\usepackage{textcomp}
\usepackage{my_chet}
\usepackage{amssymb,amsmath}

\newcommand{\cP}{\mathcal{P}}
\newcommand{\cO}{\mathcal{O}}
\newcommand{\bE}{\mathbb{E}}
\newcommand{\bR}{\mathbb{R}}
\newcommand{\be}{\begin{equation}}
\newcommand{\ee}{\end{equation}}
\newcommand{\beq}{\begin{eqnarray}}
\newcommand{\eeq}{\end{eqnarray}}
\usepackage{tikz}

\usepackage{flip-acronyms} \usepackage{tikzfeynman,contour}

\newcommand{\din}{d}
\newcommand{\dout}{d_\text{out}}

\newcommand{\aIB}{\alpha_{\text{IB}}}

\usepackage[compat=1.0.0]{tikz-feynman}

\RequirePackage[normalem]{ulem} 
\RequirePackage{color}\definecolor{RED}{rgb}{1,0,0}\definecolor{BLUE}{rgb}{0,0,1} 

\title{Neural Network Field Theories:  \\ \bigskip Non-Gaussianity, Actions, and Locality}

\author{Mehmet Demirtas$^{1,2}$\email{m.demirtas@northeastern.edu}, James Halverson$^{1,2}$\email{j.halverson@northeastern.edu}, Anindita Maiti$^{1,2,4}$\email{amaiti@perimeterinstitute.ca}, \\ Matthew D. Schwartz$^{1,3}$ \email{schwartz@g.harvard.edu}, and Keegan Stoner$^{1,2}$\email{stoner.ke@northeastern.edu}}

\affiliation{$^{1}$The NSF AI Institute for Artificial Intelligence and Fundamental Interactions\\ \vspace{.1cm}$^{2}$Department of Physics, Northeastern University,\\ Boston, MA 02115 USA\\ \vspace{.1cm}$^{3}$Department of Physics, Harvard University,\\ Cambridge, MA 02138 USA \\
\vspace{.1cm}$^{4}$School of Engineering and Applied Sciences, Harvard University,\\ Cambridge, MA 02138 USA}

\abstract{Both the path integral measure in field theory and ensembles of neural networks describe distributions over functions. When the central limit theorem can be applied in the infinite-width (infinite-$N$) limit, the ensemble of networks corresponds to a free field theory. Although an expansion in $1/N$ corresponds to interactions in the field theory, others, such as in  a small breaking of the statistical independence of network parameters, can also lead to interacting theories. These other expansions can be advantageous over the $1/N$-expansion, for example by improved behavior with respect to the universal approximation theorem. Given the connected correlators of a field theory, one can systematically reconstruct the action order-by-order in the expansion parameter, using a new Feynman diagram prescription whose vertices are the connected correlators. This method is motivated by the Edgeworth expansion and allows one to derive actions for neural network field theories. Conversely, the correspondence
allows one to engineer architectures realizing a given field theory by representing action deformations as deformations of neural network parameter densities.  As an example, $\phi^4$ theory is realized as an infinite-$N$ neural network field theory. \\


}

\begin{document}
\maketitle

\tableofcontents\newpage

\section{Introduction}

The last decade has seen remarkable progress in machine learning (ML) in a wide variety of fields, including traditional ML fields such as natural language processing, image recognition, and gameplay (see \cite{LeCun2015, GoodBengCour16} for reviews, and \cite{NIPS2017_3f5ee243, Silver2017} for some breakthroughs in the literature), but also in the physical sciences \cite{RevModPhys.91.045002}, and more recently to obtain rigorous results in pure mathematics \cite{Carifio, Gukov_2021,Davies2021,gukov2023searching}. This progress has been facilitated in part by the increasing complexity of deep neural networks, both in terms of the number of parameters appearing in them and their architecture. However, despite their empirical success, the theoretical foundations of deep neural networks are still not fully understood.
Natural questions emerge:
\begin{itemize}
    \item Are ideas from the sciences, such as physics, useful in neural network theory?
    \item As it develops, does ML theory lead to progress in the sciences?
\end{itemize}
A growing literature (see below), gives an affirmative answer to the first, but the second is less clear; 
it is applied ML, not theoretical ML, that is primarily used in the sciences.

In this paper we explore both of these questions by further developing a correspondence between neural networks and field theory. This connection was already implicit in Neal's Ph.D. thesis \cite{neal} in the $1990$'s, where he demonstrated that an infinite width single-layer neural network is (under appropriate assumptions) a draw from a Gaussian process (GP). This is the so-called neural network / Gaussian process (NNGP) correspondence, and in recent years it has been shown that most modern NN architectures \cite{Matthews2018GaussianPB,Novak2018BayesianCN,GarrigaAlonso2019DeepCN,yangTPorig,yangTP1,yangTP2} have a parameter $N$ such that the NN is drawn from a GP in the $N\to \infty$ limit. The NNGP correspondence is of interest from a physics perspective because Gaussian processes are generalized non-interacting (free) field theories, and neural networks provide a novel way to realize them. Non-Gaussianities emerge at finite-$N$, which correspond to turning on interactions that are generally non-local, and may be captured by statistical cumulant functions, known as connected correlators in physics. As we will see, since Gaussianity in the $N\to \infty$ limit emerges by the Central Limit Theorem (CLT), non-Gaussianities may be studied more generally by parametrically violating necessary conditions of the CLT.

These results provide a first glimpse that there is a more general NN-FT correspondence that should be developed in its own right, taking inspiration from both physics and ML. In this introduction we will review the central ideas of the correspondence and introduce principles for understanding the literature, which we review in part. Readers familiar with the background are directed to Section \ref{subsec:intro-results} for a summary of our results.

\subsection{NN-FT Correspondence}

At  first glance, neural networks and field theories seem very different from one another.
However, in both cases, the central objects of study are \textit{random functions}. The random function $\phi$ associated to a neural network is defined by its \emph{architecture}, which is a composition of simpler functions that involves parameters $\theta$. At program initialization, parameters are drawn as $\theta \sim P(\theta)$, yielding a randomly initialized neural network, i.e. a random function.
In field theory, the random functions are simply the fields themselves, typically described by specifying their probability density function directly, $P(\phi) = \exp(-S[\phi])$, via the Euclidean action functional $S[\phi]$; we work in Euclidean signature throughout.

\begin{figure}[t]
\centering
\begin{tikzpicture}[scale=.75]
  \coordinate (A) at (0,0);
  \coordinate (B) at (5,0);
  \coordinate (Au) at (.2,1.2);
  \coordinate (Bu) at (4.8,1.2);
  \coordinate (Ad) at (.2,-1.2);
  \coordinate (Bd) at (4.8,-1.2);
  
  \draw (A) circle (1) node {\Large NN};
  \draw (B) circle (1) node {\Large FT};
  
  \draw[->,bend left=30] (Au) to node[above] {} (Bu);
  \draw[->,bend right=330] (Bd) to node[below] {} (Ad);
\end{tikzpicture}
\label{fig:nn-ft}
\caption{In a NN-FT correspondence, ideas from one may give insights into the other. In this paper we are primarily interested in understanding when neural network field theories exhibit physical principles such as non-Gaussianity and locality, with an eye towards applications in both ML and especially physics in the future.}
\end{figure}
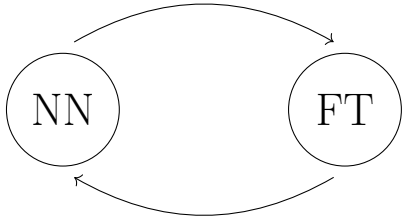

We therefore have two different origins for the statistics of a field theory. To exemplify the point, consider a field theory defined by an ensemble of networks or fields $\phi: \bR \to \bR,$
\be 
 \phi(x) = a \, \sigma(b \, \sigma(c \, x)) \qquad \qquad a \sim P(a), \,\, b \sim P(b), \,\, c \sim P(c), \label{eqn:first_arch}
\ee 
where $\sigma: \bR \to \bR$ acts element-wise and is generally taken to be non-linear.
Here the statistics of the ensemble arise from how it is \emph{constructed}, rather than from the density $\exp(-S[\phi])$ over functions from which it is drawn. We will refer to such a description as the \emph{parameter space} description of a \emph{neural network field theory}. The construction of $\phi$ defined in \eqref{eqn:first_arch} has two parts, the architecture that defines its functional form, and the choice of distributions from which the parameters $a$, $b$, and $c$ are drawn. This particular architecture is a feedforward network with depth two, width one, and activation function $\sigma$.
In this description of the field theory, one does not necessarily know the action $S[\phi]$, but the theory may nevertheless be studied because the architecture and parameter densities define its statistics. 

For instance, the correlation functions of a neural network field theory can be expressed as 
\be 
G^{(n)}(x_1, \dots, x_n) := \mathbb{E}[\phi(x_1)\dots\phi(x_n)] = \int d\theta \, P(\theta)\, \phi(x_1)\dots\phi(x_n)\label{eqn:psG},
\ee
where we denote the set of parameters of the neural network by $\theta$, and the network / field $\phi$ depends on parameters through its architecture. Alternatively, we could provide a \emph{function space} description of the theory by specifying the action $S[\phi]$ and express the correlation functions as
\be \label{eqn:fsG}
G^{(n)}(x_1, \dots, x_n) = \int D\phi \, e^{-S[\phi]} \, \phi(x_1)\dots\phi(x_n),
\ee
as in a first course on quantum field theory. These expressions may be derived from the partition function
 \be 
 Z[J] = \bE[e^{\int d^dx J(x) \phi(x)}]\label{eqn:Z},
 \ee 
 where the parameter space and function space results arise by specifying how the expectation value is computed, 
 \begin{align}
 Z[J] &= \int d\theta  \, P(\theta)\, e^{\int d^dx J(x) \phi(x)} \label{eqn:psZ}\\
 Z[J] &= \int D\phi \, e^{-S[\phi]+\int d^dx J(x) \phi(x)}. \label{eqn:fsZ}
 \end{align}
 In this work, many calculations will be carried out in terms of a general expectation value $\bE[\cdot]$ that denotes agnosticism towards the origin of the statistics; explicit calculations may be carried out by replacing $\bE$ with one description or the other, as in passing from a general expression \eqref{eqn:Z} to those of parameter space \eqref{eqn:psZ} and function space \eqref{eqn:fsZ}.
 
Parameter space and function space provide two different descriptions of a field theory, which could be thought of as different duality frames \cite{maiti2021symmetryviaduality}. When one defines a field theory by a neural network architecture, the parameter space description is readily available, but the action is not known, a priori. However, if the parameter distributions are easy to sample then the fields are also easy to sample: one just initializes neural networks on the computer. On the other hand, in field theory we normally proceed by first specifying an action; in this case, the probability of a given field configuration is known because $P[\phi] = \exp(-S[\phi])$ is known, but fields are notoriously hard to sample, as evidenced by the proliferation of Monte Carlo techniques in lattice field theory.

\bigskip
\noindent\emph{Example: NNGP Correspondence in Parameter Space and Function Space}

Let us study an example to make the abstract notions more concrete. Consider a fully-connected feedforward network $\phi: \bR^d \to \bR$ with depth one and width $N$,
\be 
\phi(x) = \sum_{i=1}^N \sum_{j=1}^d a_i \,\sigma\, (b_{ij} x_j), \qquad\qquad a\sim \mathcal{N}(0,\sigma^2/N),\,\, b\sim \mathcal{N}(0,\sigma^2/d),
\ee 
where $\sigma$ is an elementwise non-linearity such as $\text{tanh}$ or $\text{ReLU}(z):= \text{max}(0,z)$. Here, the set of parameters $\theta$ is given by the union of the $a$-parameters and the $b$-parameters. As we will see in detail in Section \ref{sec:NG}, if the parameters are drawn independently then the connected correlation functions 
\be 
G^{(2k)}_{c}(x_1,\dots,x_{2k}) \propto \frac{1}{N^{k-1}},
\ee 
and the odd-point correlation functions vanish due to $a$ having zero mean. In the $N\to \infty$ limit, also known as the Gaussian Process (GP) limit, then, the only non-vanishing connected correlator has two points,
\be 
G^{(2)}_{c}(x_1,x_2)
\ee 
which demonstrates that the theory is Gaussian; this is the NNGP correspondence. Concretely, following \cite{williams}, we may compute the two-point function as 
\be 
G^{(2)}(x, y) = \bE[\phi(x)\phi(y)] = \int \, da \, db\, P(a) P(b) \,\,a_{i_1} \sigma( b_{i_1j_1} x_{j_1}) \,\, a_{i_2} \sigma( b_{i_2j_2} y_{j_2})
\ee 
where we have used Einstein summation and left the details of the Gaussian parameter densities $P(a)$ and $P(b)$ implicit. For a fixed choice of $\sigma$ one may evaluate this integral analytically or via Monte Carlo sampling, resulting in the two-point function; analytic integrated results for $\sigma = \text{tanh}$ and $\sigma=\text{Erf}$ are presented in \cite{williams}.
Since the parameter space calculation establishes Gaussianity of the theory, we infer the action
\be 
S[\phi] = \int d^dx\, d^dy \, \phi(x) \, G^{(2)}(x,y)^{-1}\, \phi(y), \label{eqn:basic}
\ee
where the inverse of the two-point function satisfies $\int d^d y \,G^{(2)}(x,y)^{-1} G^{(2)}(y,z) = \delta^{(d)}(x-z)$. \

As a concrete example, we refer the reader to Section \ref{sec:phi4}, which recalls a neural network realization of free scalar field theory from \cite{halverson2021building} that uses a $\cos$ activation. In that case we have 
\begin{equation}
G^{(2)}(x,y)^{-1} = \delta(x-y) (\nabla^2+m^2)
\end{equation}
which reproduces the usual free scalar action
\begin{equation}
S[\phi] = \int d^d x\, \phi(x)\left(\nabla^2+m^2\right)\phi(x),
\end{equation}
in this case realized via a concrete neural network architecture.

Thus, in the GP limit, both the parameter space and function space descriptions of the field theory are readily available. Building on \cite{Naveh_2021} using the Edgeworth expansion, we will see methods for computing approximate actions at finite-$N$, and we will also develop techniques to engineer desired actions.

\subsection{Organizing Principles and Related Work}

We have discussed a foundational principle underlying the NN-FT correspondence, that parameter space and function space provide two different descriptions of the statistics of an ensemble of neural networks or fields. Though we have given an example, and there are many more, we are still in very general territory and it is not clear where to go. Accordingly, we would like to provide other organizing principles: 
\begin{itemize}
\item \textbf{NN-for-FT vs. FT-for-NN:} are we aiming to better understand physics or ML?
\item \textbf{Fixed Initialization vs. Learning:}  are we aiming to understand a fixed NN-FT at initialization, or a one-parameter family of NN-FTs defined by some dynamics, such as ML training dynamics or field theory flows?
\end{itemize}
Much of the existing literature can be classified within each of these principles, and they also set context for discussing our results.  We will first review some results for network ensembles at initialization, and then during and after training. With these ideas in place, we will turn to the idea of using NN-FT in service of field theory.

\bigskip For literature that is most similar in perspective to this introduction (prior to this reference section), see \cite{halverson2021building} and the works that preceded it \cite{Halverson_2021,maiti2021symmetryviaduality}, by subsets of the authors.

\bigskip 
\noindent \textbf{Initialization.} A neural network with parameters $\theta$ and parameter distribution $P(\theta)$ is initialized on a computer by drawing $\theta \sim P(\theta)$ and inserting them into the architecture, generating a random function $\phi(x)$ that is sampled from a distribution $P(\phi)$ that may or may not be known. In the $N\to \infty$ NNGP limit, $P(\phi)$ is Gaussian. This was shown for feed forward networks in Neal's thesis \cite{neal}, as well as more recently in \cite{Matthews2018GaussianPB, Novak2018BayesianCN,GarrigaAlonso2019DeepCN}; was generalized to a plethora of architectures, e.g. convolutional layers \cite{yangTPorig,yangTP1, yangTP2, conv16, Fukushima1980NeocognitronAS, rumelhart1985learning, lecun1998gradient, lecun1999object}, recurrent layers, graph convolutions \cite{bruna2013spectral, henaff2015deep,Duvenaud2015MolFingerPrintConvolutionNetwork, li2015gated, defferrard2016convolutional, kipf2016semi}, skip connections \cite{skip1, skip2}, attention \cite{bahdanau2014neural, vaswani2017attention}, and batch  /layer normalization in \cite{ioffe2015batch, ba2016layer}, pooling \cite{lecun1998gradient, lecun1999object}, and transformers \cite{hron2020infinite, dinan2023effective}. The generality of this result arises from the generality in which central limit theorem behavior manifests itself in neural networks; see \cite{yangTPorig,yangTP1, yangTP2} for a systematic treatment in the tensor programs formalism.

Since Gaussianity follows from the central limit theorem, one generally expects non-Gaussianities in the form of $1/N$-corrections. Study of these non-Gaussianities was initiated a few years ago; e.g., \cite{Yaida2019NonGaussianPA} computed leading non-Gaussianities via the connected four-point function, \cite{antognini2019finite} showed for deep feedforward networks how $P(\phi)$ is perturbed by $1/N$-corrections, \cite{Halverson_2021} proposed using effective field theory to model non-Gaussian $P(\phi)$ for neural networks, and \cite{Roberts_2022} developed an effective theory approach and an $L/N$ expansion that controls feature learning in deep feedforward networks; for concreteness in our examples, we are interested in the distribution of networks at initialization and take $L=1$.
This $L/N$ expansion allowed \cite{Roberts_2022} to also study signal propagation through the network, identify universality classes, and tune hyperparameters to criticality.  

Methods borrowed from field theory have been useful in studying NNs at initialization. For example, perturbative methods like Feynman diagrams were employed in \cite{Dyer2020AsymptoticsOW, Halverson_2021,erdmenger2021quantifying, grosvenor2022edge}. Various schemes for renormalization group flow, including non-perturbative ones, were applied to NNs in   \cite{Erbin_2022, erbin2022renormalization}.  Global symmetries of NN-FTs were shown to arise from symmetry invariances of NN parameter distributions in \cite{maiti2021symmetryviaduality}. While the results of this paper were being finalized, a recent paper \cite{banta2023structures} brought forward a different diagrammatic approach to effective field theories in deep feedforward networks.

\bigskip 
\noindent \textbf{Learning.} 
Although we do not study the dynamics of learning in this paper, it is a goal for future work. Therefore, we would like to review some of the literature.

Neural networks may be trained to perform useful tasks via a variety of learning schemes, such as supervised learning or reinforcement learning, that utilizes a learning algorithm to update the system, such as stochastic gradient descent. In practice this involves training one or a handful of randomly initialized neural networks to convergence. However, in general there is nothing special about the initial networks that were trained; in the absence of compute limitations, one would prefer to train \emph{all} the networks and compute an ensemble average at convergence. Theoretically, this amounts to tracking the distributional flow of the neural network ensemble, and in principle it may be done in either parameter space or function space.

In the $N\to \infty$ limit, most known architectures define neural networks that are draws from Gaussian processes. Since the architecture defines a GP, it could be used as a prior in Bayesian inference, the learning algorithm of interest in Neal's original work \cite{neal}. On the other hand, gradient descent with continuous time is governed by the neural tangent kernel (NTK) \cite{Jacot2018NeuralTK},
which becomes deterministic and training time $t$-independent in the so-called frozen-NTK limit. In this limit, $N \to \infty$ and the neural network dynamics is well-approximated by that of a model that is linear in the neural network parameters. This frozen behavior is a vast simplification of the dynamics and is known to exist for many architectures, such as convolutional neural networks\cite{arora2019exact}, graph neural networks\cite{du2019graph}, recurrent networks\cite{alemohammad2021recurrent,alemohammad2021enhanced}, and attention layers\cite{hron2020infinite}. For supervised learning with MSE loss, the neural network ensemble trained under gradient descent remains a GP for all times $t$, including $t\to \infty$, with known mean and covariance; the dynamics becomes that of kernel regression, with kernel given by the frozen-NTK. How is this related to Neal's desire to relate Bayesian inference and trained neural networks? If all but the last layer's weights are frozen, then the NTK is the NNGP kernel and the distribution of the neural network ensemble converges to the GP Bayesian posterior as $t\to\infty$.

In summary, in the $N\to \infty$ limit, the distribution of the neural network ensemble is Gaussian. If it undergoes supervised training with MSE loss, it remains Gaussian at all times and converges to the Bayesian GP posterior in a particular case \cite{Lee2019WideNN}. In general, however, gradient descent induces non-Gaussianities.

At finite-$N$, the neural network ensemble is non-Gaussian. In the Bayesian context, this defines a non-Gaussian prior, and inference may be performed for weakly non-Gaussian priors via a $1/N$-expansion \cite{Yaida2019NonGaussianPA}. In the gradient descent context, the NTK is no longer frozen and evolves during training, significantly complicating the dynamics. Work by Roberts, Yaida, and Hanin develops a theory of an evolving NTK in \cite{Roberts_2022}. They apply it in detail to fully-connected networks of depth $L$, demonstrate the relevance of $L/N$ as an expansion parameter, and develop an effective model for the dynamics. Such $1/N$ corrections to dynamical NTK were previously studied by other authors in \cite{pmlr-v119-huang20l, aitken2020asymptotics, Dyer2020AsymptoticsOW}. Pehlevan et. al. have developed a systematic understanding of the evolution of NTK and parametric interpolations between rich and lazy training regimes using the framework of dynamical mean field theory, see \cite{bordelon2023dynamics}. Some of these authors have studied the $O(1/N)$ suppressed corrections to training dynamics of finite width Bayesian NNs in \cite{NEURIPS2021_cf9dc5e4}. A separate work, \cite{seroussi2022separation}, presents close-to-Gaussian NN processes including stationary Bayesian posteriors in the joint limit of large width and large data set, using $1/N$ as an expansion parameter. Moreover, the authors of \cite{krippendorf2022duality} explore a correspondence between learning dynamics in the continuous time limit and early Universe cosmology, and  \cite{Fischer_2022, dick2023linking, Huang_2018} analyzes connected correlation functions propagating through neural networks.

\bigskip
\noindent \textbf{NN-for-FT.} 

Neural networks, including the ones we have discussed thus far, generally have $\bR^n$ as their domain and therefore naturally live in Euclidean signature. They define statistical field theories that may or may not have analytic continuations to 
quantum field theories in Lorentzian signature. Nevertheless, statistical field theories are interesting in their own right and NN-FT provides a novel way to study them.

Using an architecture to define a field theory enables a parameter space description that makes sampling, and therefore numerical simulation on a lattice, easy. If one can determine an easily sampled NN architecture that engineers standard Euclidean $\phi^4$ theory, for instance, this could lead to improved results on the lattice by avoiding Monte Carlo entirely \footnote{This lattice approach should be contrasted with works \cite{Albergo_2019, Abbott_2022, gerdes2022learning} that train a normalizing flow to give proposals for the accept / reject step of MCMC.} This is an engineering problem that is work-in-progress; it is not clear that the $\phi^4$ NN-FT realization in this work is easily sampled. Alternatively, by simply fixing an easily sampled architecture with interesting physical properties such as symmetries and strong coupling, lattice simulation could be performed immediately.

\bigskip For uses in fundamental and formal quantum physics, one might wish to know when a neural network architecture defines a \emph{quantum} field theory (QFT). Since NN architectures are usually defined in Euclidean signature, we may instead ask when a Euclidean field theory admits an analytic continuation to Lorentzian signature that defines a QFT. The situation is complicated by the fact that in general we do not know the action, but instead have access to the Euclidean correlation functions, expressed in parameter space. 

Fortunately, the Osterwalder-Schrader (OS) theorem \cite{Osterwalder1973}  of axiomatic field theory gives necessary and sufficient conditions, expressed in terms of the correlators, for the existence of a QFT after continuation. The axioms include
\begin{itemize}
\item \textbf{Euclidean Invariance.} Correlation functions must be Euclidean invariant, which becomes Lorentz invariance after analytic continuation. See \cite{halverson2021building} for an infinite ensemble of NN architectures realizing Euclidean invariance.
\item \textbf{Permutation Symmetry}. Correlation functions must be invariant under permutations of their arguments, a collection of points in Euclidean space. This is automatic in NN-FTs with scalar outputs.
\item \textbf{Reflection Positivity}. Correlation functions must satisfy a positivity condition known as reflection positivity, which is necessary for unitarity and the absence of negative-norm states in the analytically continued theory.
\item \textbf{Cluster Decomposition.} Correlation functions must satisfy cluster decomposition, which says that interactions must shut off at infinite distance. As a condition on connected correlators, cluster decomposition is 
\begin{equation}
\lim_{b\to \infty} \,\, G^{(n)}_{c}(x_1,\dots,x_p, x_{p+1} + b, \cdots, x_n + b) \to 0,
\end{equation}
for any value of $1< p < n$. We have assumed permutation symmetry to simplify notation, putting the shifts into $x_{p+1}$ into $x_n$.
\end{itemize}

These ideas were utilized in \cite{halverson2021building} to define neural network \emph{quantum} field theories: a NN-QFT is a neural network architecture whose correlation functions satisfy the OS axioms, and therefore defines a QFT upon analytic continuation. To date, the only known example is a NN architecture that engineers a standard free scalar field theory in $d$-dimensions, though we improve the situation in this work by developing techniques to engineer local Lagrangians, which automatically satisfy the OS axioms.  To make further progress on NN-QFT in a general setting, one needs especially a deeper understanding of reflection positivity and cluster decomposition in interacting NN-FTs; we study the latter.

\subsection{Summary of Results and Paper Organization}\label{subsec:intro-results}

Since there are a number of different themes and concepts in this paper, we would like to highlight some of the major conceptual results:
\begin{itemize}
    \item \textbf{Parametric Non-Gaussianity: $1/N$ and Independence Breaking.} \\ Section \ref{sec:NG} approaches interactions in NN-FT (non-Gaussianity) by parametrically breaking necessary conditions for the central limit theorem to hold. Violating the infinite-$N$ limit is well studied, but we also systematically study interactions arising from the breaking of statistical independence, and apply these ideas in examples.
    \item \textbf{Computing Actions with Feynman Diagrams.} \\  In Section \ref{sec:actions} we develop a general field theory technique for computing the action diagrammatically. The coupling functions are computed with a new type of connected Feynman diagram, whose vertices are the connected correlators. This is a swapping of the normal role of couplings and connected correlators, which arises from a ``duality" that becomes apparent via the Edgeworth expansion. The technique is also applied to NN-FT, including an analysis of how actions may be computed in the two regimes of parameteric non-Gaussianity developed in Section \ref{sec:NG}, $1/N$ and independence breaking.
    \item \textbf{Engineering Actions in NN-FT.}  \\ In Section \ref{sec:designing actions} we develop techniques for engineering actions in NN-FT. This is to be distinguished from the approach of Section \ref{sec:actions}: instead of fixing an architecture, computing its correlators, and then computing its action via Feynman diagrams, in Section \ref{sec:designing actions} we fix a desired action and develop techniques for designing architectures that realize the action. Adding a desired term to the action manifests itself in NN-FT by deforming the parameter distribution, which breaks statistical independence if it is a non-Gaussianity. Using this technique, local actions may be engineered at infinite-$N$.
    \item \textbf{$\phi^4$ as a NN-FT.} \\ In Section \ref{sec:phi4} we design an infinite width neural network architecture that realizes $\phi^4$ theory, using the techniques that we developed.
    \item \textbf{The Importance of $N\to \infty$ for Interacting Theories.} In physics, interesting theories defined by a fixed action $S$ generally have a wide variety of finite action field configurations, which have non-zero probability density. This is potentially at odds with the universal approximation theorem: if a single finite-action configuration cannot be realized by an architecture $A$, but only approximated, then any NN-FT associated to $A$ cannot realize the field theory associated to $S$. If the $1/N$ is an expansion parameter for both non-Gaussianities and the degree of approximation, as e.g. with single-layer width-$N$ networks, this simple no-go theorem suggests that exact NN-FT engineering of well-studied theories in physics occurs most naturally at infinite-$N$, as we saw in the case of  $\phi^4$ theory.
\end{itemize}
These are highlights of the paper. For more detailed summaries of results, we direct you to the beginning of each section.

\section{Connected Correlators and the Central Limit Theorem \label{sec:NG}}

Interacting field theories with a Lagrangian description are defined by non-Gaussian field densities $\exp(-S[\phi])$. If the non-Gaussianities are small, the theory is close to Gaussian and weakly interacting, in which case correlation functions may be computed in perturbation theory using Feynman diagrams. The non-Gaussianities are captured by the higher connected correlation functions, which vanish in the Gaussian limit. They are known as cumulants in the statistics literature and may be obtained from a generating functions $W[J]$ as  
\begin{equation}
G^{(n)}_{c}(x_1,\dots, x_n) := \left(\frac{\delta}{\delta J(x_1)} \dots \frac{\delta}{\delta J(x_n)} W[J]\right)\Bigg|_{J=0}, \qquad \qquad W[J] := \ln Z[J].
\end{equation}
In the absence of a known Lagrangian description, connected correlators still encode the presence of non-Gaussianities, since the theory is Gaussian if $G^{(n)}_c=0$ for $n>2$.

In this section we systematically study non-Gaussianities in NN-FT. Since the parameter space description exists for any NN-FT, we choose to study non-Gaussianities via connected correlators (rather than actions), which may be studied in parameter space even when the action is unknown. We are interested in non-Gaussianities in NN-FT for a number of reasons. In the NN-for-FT direction, it is important for understanding interactions in the associated field theories. Conversely, in the FT-for-NN direction, understanding non-Gaussianities is important for capturing the statistics of finite networks and networks with correlations in the parameter distributions, which generally develop during training. 

\bigskip
The essential idea in our approach is to recall the origin of Gaussianity, and then parametrically move away from it. 
Specifically,
many field theories defined by neural network architectures admit an $N\to \infty$ limit in which they are Gaussian, and the Gaussianity has a statistical origin: the Central Limit Theorem (CLT). The CLT states that the distribution of the standardized sum of $N$ independent and identically distributed random variables approaches a Gaussian distribution in the limit $N \to \infty$.  Therefore we may systematically study non-Gaussianities in neural network field theories by violating assumptions of the CLT, e.g. via $1/N$ corrections and breaking the independence condition, both of which affect connected correlators.

There are a number of results and themes in this section, which is organized as follows:
\begin{itemize}
\item \textbf{Central Limit Theorem.} In Section \ref{sec:CLT} we review the CLT from the perspective of cumulant generating functionals, which will be useful in NN-FT since in general we do not have a simple expression for the action but do have access to cumulants. 
\item \textbf{Independence Breaking.} In Section \ref{sec:independence_breaking} we introduce how non-Gaussianities may also arise by violating the statistical independence assumption of the CLT. We characterize this by a family of joint densities with parameter $\alpha$ that factorize (become independent) when $\alpha=0$. We study the $\alpha$-dependence of cumulants via Taylor series, showing that $\alpha$ controls non-Gaussianities independently of those arising from $1/N$-corrections. A simple example of independence-breaking induced non-Gaussianities at $N=\infty$ is given in Section \ref{sec:independence_breaking_example_inf_N}.
\item \textbf{Connected Correlators and Interactions in NN-FT.} In Section \ref{sec:NG_continuum} we study non-Gaussianities in NN-FT, decomposing the field $\phi(x)$ into $N$ constituent neurons as in \cite{halverson2021building}. We study the case of independent neurons in Section \ref{sec:continuumfiniteN}, where we present the $N$-scaling of connected correlators and also two examples: single-layer Cos-net, which exhibits full Euclidean symmetry in all of its correlators, and $d=1$ ReLU-net, which we show exhibits an interesting bilocal structure in its two-point and four-point functions. 

In Section \ref{sec:continuum_ib} we turn to breaking neuron independence in NN-FT,  building on the independence breaking results of \cite{halverson2021building}, which gives a new source of interactions and a generalized formula for connected correlators. Specifically, we introduce a general formalism for the expansion of the cumulant generating functional in terms of independence-breaking parameters, and therefore the computation of connected correlators. As an example, we deform the Cos-net theory to have non-independent neurons via non-independent input weights, doing the deformation in a way that preserves Euclidean invariance, and compute the independence-breaking correction to the connected four-point function.
\item \textbf{Identical-ness Breaking.} Interactions may also arise from breaking the identical-ness assumption of the CLT. See Appendix \ref{appendix_eg} for an example of a NN-FT with non-Gaussianities arising from identical-ness breaking.
\end{itemize}
Equipped with two different types of parameters that induce non-Gaussianity in connected correlators, $1/N$ and independence-breaking parameters, we will see how this may be used to approximate actions in Section \ref{sec:actions}.

\subsection{Review: Central Limit Theorem from Generating Functions\label{sec:CLT}}

In order to understand non-Gaussianities in NN-FTs, it is useful to recall essential aspects of the Central Limit Theorem in the case of a single random variable, since they carry over to the NN-FT case. We will do so using the language of generating functions and cumulants (connected correlators), since we may use them to study Gaussianity and non-Gaussianity even if the NN-FT action is unknown.

\medskip
Of course, the CLT is among the most fundamental theorems of statistics. There are many variants of it in the literature, with different sets of assumptions. Here, we will describe a particularly simple version of it and provide a proof, showing how key assumptions come into play. For a more in depth discussion of the CLT, see e.g. \cite{Dedecker2007}.

Consider $N$ random variables $X_i$. Assume that they are \emph{identical}, \emph{independent}, \emph{mean-free}, and have \emph{finite variance}. The CLT states that the standardized sum 
\begin{equation}
\label{eqn:phi_CLT}
\phi = \frac{1}{\sqrt{N}} \sum_{i=1}^N X_i
\end{equation}
is drawn from a Gaussian distribution in the limit $N \to \infty$. In other words, even if the $X_i$ are sampled from complicated, non-Gaussian distributions, these details wash out and their sum is drawn from a Gaussian distribution. 

To see the Gaussianity in a way that may be extrapolated to NN-FT, it is useful to introduce generating functions. The moment generating function of $\phi$ is defined as
\begin{equation}
    Z_\phi[J] := \bE[e^{J \phi}]  =  \bE[e^{J\sum_i X_i / \sqrt{N}}],
\end{equation}
from which we can extract the moments by taking derivatives,
\begin{equation}
    \mu^\phi_r := \bE[\phi^r] = \Bigl(
    \frac{\partial}{\partial J} \Bigr)^r Z_\phi[J].
\end{equation}
In physics language, $J$ is the source, $Z_\phi[J]$ is the partition function, and $\mu^\phi_r$ is the $r^\text{th}$ correlator of $\phi$.
The cumulant generating functional (CGF) of $\phi$ is the logarithm of the moment generating functional
\begin{equation} \label{eqn:singlevarcgf}
W_\phi[J] := \log \,\bE[e^{J \phi}]  = \log \,\bE[e^{J\sum_i X_i / \sqrt{N}}] ,
\end{equation}
and the cumulants $\kappa^\phi_r$ are computed by taking derivatives of $W_\phi[J]$,
\begin{equation}
    \kappa^\phi_r := \Bigl(
    \frac{\partial}{\partial J} \Bigr)^r W_\phi[J].
\end{equation}
A random variable is Gaussian only if its cumulants $\kappa^\phi_{r>2}$ vanish. Fundamental properties of CGFs include 
\begin{align}
W_{X+c}[J] = c J + W_{X}[J], \\
W_{c X}[J] = \log \, \bE[e^{J \, c x}] = W_{X}[c J]
\end{align}
where $c \in \mathbb{R}$ is a constant, which imply 
\begin{align}
\kappa^{X+c}_1 = \kappa^X_1 + c \qquad \kappa^{X+c}_{r>1} = \kappa^X_{r>1}\\
\kappa^{cX}_r = c^r \,\kappa^X_r,
\label{eqn:cumiden}
\end{align}
respectively. 

We would like to see the Gaussianity of $\phi$ under CLT assumptions by computing cumulants. This is possible since $\kappa_{r>2}=0$ is necessary for Gaussianity; conversely, we may study non-Gaussianities in terms of non-vanishing higher cumulants.
Specifically, for a sum of \emph{independent} random variables the moment generating function factorizes,
\begin{equation}
    Z_{X_1+\dots+ X_N}[J] = \prod_i^N Z_{X_i}[J].
\end{equation}
Consequently, the CGF and the cumulants become
\begin{align}
W_{X_1+\dots+ X_N}[J] = W_{X_1}[J] + \dots + W_{X_N}[J], \\
\kappa_r^{X_1+\dots+X_N} = \kappa_r^{X_1} + \dots + \kappa_r^{X_N} \label{eqn:phisumcumulants_ind}.
\end{align}
Using the identities in $\eqref{eqn:cumiden}$ we can write the cumulants of $\phi$ as
\begin{equation}
\kappa_r^\phi = \frac{\kappa_r^{X_1} + \dots + \kappa_r^{X_N}}{N^{r/2}}.
\end{equation}
When the $X_i$ are \emph{identical} this simplifies to
\begin{equation}
    \kappa_r^\phi = \frac{\kappa^{X_i}_r}{N^{r/2-1}}.
\label{eqn:phicumulants_N_scaling}
\end{equation}
The cumulants $\kappa^\phi_{r>2}$ vanish in the $N\to \infty$ limit. To establish that $\phi$ is Gaussian, we also need to show that $\kappa^\phi_1$ and $\kappa^\phi_2$ are finite.
As the $X_i$ are mean-free, $\kappa_1^\phi = \kappa_1^{X_i}/\sqrt{N}=0$, while $\kappa_2^\phi = \kappa_2^{X_i}$ is finite by assumption. Thus, $\phi$ is Gaussian distributed. This is the Central Limit Theorem, cast into the language of cumulants. 

We emphasize that this result relies not only on the $N\to \infty$ limit, but also on the independence assumption \eqref{eqn:phisumcumulants_ind}.

\subsection{Non-Gaussianity from Independence Breaking} \label{sec:independence_breaking}
We wish to study the emergence of non-Gaussianity by breaking the independence condition.

To do so, we must parameterize the breaking of statistical independence. Let $p(X;\alpha)$ be a family of joint distributions on $X_i$ parameterized by a hyperparameter $\alpha$ that must be chosen in order to define the problem. We choose the family of joint distributions to be of the form
\begin{equation}
p(X;\alpha=0)= \prod_i p(X_i),
\end{equation}
i.e. $p(X)$ is independent in the $\alpha \to 0$ limit, but $\alpha \neq 0$ in general controls the breaking of independence. Then we obtain
\begin{equation}
    W_{\phi}[J] = \log \bE [e^{J\sum_i X_i/\sqrt{N}}] = \log \int \prod_j dX_j \, p(X;\alpha)\,  e^{J\sum_i X_i/\sqrt{N}}
\end{equation}
which when expanded around $\alpha=0$ yields 
\begin{equation} 
    \label{eqn:phiCGF_simple_intermediate}
    W_{\phi}[J] = \log \left[\prod_j \bE_{p(X, \alpha=0)}[e^{JX_j/\sqrt{N}}] + \sum_{k=1}^\infty \frac{\alpha^k}{k!}\, \int \prod_j dX_j \,\,  e^{J\sum_i X_i/\sqrt{N}} \,\,\partial_\alpha^k p(X;\alpha)|_{\alpha=0} \right],
\end{equation}
where the first term of the log uses independence of $p(X;\alpha=0)$. 

To deal with the $\alpha$-dependent terms, we generalize a trick appearing regularly in machine learning, e.g. in the policy gradient theorem in reinforcement learning. There, the fact that $ p \, \partial_\alpha \log p = \partial_\alpha p$ allows us to write
\be 
\partial_\alpha \bE[\mathcal{O}] = \bE[\cO \, \partial_\alpha \log p]
\ee 
for any $\alpha$-independent operator $\mathcal{O}$.
Generalizing, we define
\be
\cP_k := \frac{1}{p} \partial^k_\alpha p,
\ee
and note that it satisfies the recursion relation
\be 
\cP_{k+1} = \cP_1 \cP_k + \partial_\alpha \cP_k,
\ee
which allows for efficient computation. We can then write \eqref{eqn:phiCGF_simple_intermediate} as
\be
W_{\phi}[J] = \log \left[\prod_j \bE_{p(X, \alpha=0)}\Big[e^{JX_j/\sqrt{N}}\Big] + \sum_{k=1}^\infty \frac{\alpha^k}{k!}\, \bE_{p(X,\alpha=0)}\Big[e^{J\sum_i X_i/\sqrt{N}} \,\cP_k|_{\alpha=0}\Big] \right].
\ee
In the limit $\alpha \to 0$, the $X_j$ become independent, and we have
\be 
\lim_{\alpha \to 0}W_{\phi}[J] = \sum_j \log \bE_{p(X_j)}\left[e^{JX_j/\sqrt{N}} \right] = \sum_j \lim_{\alpha \to 0}W_{X_j/\sqrt{N}}[J],
\ee
where $\phi$ is now a sum of $N$ independent variables $X_j$, and its CGF is the sum of CGFs of $X_j/\sqrt{N}$, as expected; details of the calculations are in Appendix \eqref{app:CGFedgeworth}.

\bigskip 
We have now discussed two mechanisms that result in non-Gaussianities: $1/N$ corrections and independence breaking. While one can use either or both of these mechanisms to generate and control non-Gaussianities, more caution is required to use independence breaking alone, at infinite $N$. This is because the non-Gaussianities that are generated by independence breaking might depend on $N$ as well as $\alpha$. For example, if the leading corrections to higher cumulants $\kappa^\phi_{r}$ scale as $\alpha N^{a_r}$ with $a_r<0$ for all $ r>2$, $\phi$ will be Gaussian regardless of independence breaking. While if $a_r>0$, $\kappa^\phi_{r}$ will diverge, which is  undesirable.
In the following, we will 
present an example where $a_r=0$ for all $r$ and the non-Gaussianities are generated by independence breaking alone.

\subsubsection*{Example: Independence Breaking at Infinite N \label{sec:independence_breaking_example_inf_N}}

Let us provide an example of independence breaking non-Gaussianities that persist in the $N\to \infty$ limit, showing how one can control higher cumulants by adjusting the correlations between random variables.
Consider the normalized sum of $N$ random variables,
\begin{equation}
    \phi = \frac{1}{\sqrt{N}} \sum_{i=1}^N X_i,  
\end{equation} 
where $X_i$ is the product of two random variables $a_i$ and $h_i$,
\be
    X_i = a_i h_i.
\ee
This architecture can be interpreted as the last layer of a fully connected neural network, where $h_i$ are the outputs of the neurons in the previous layer, $a_i/\sqrt{N}$ are the weights, and $\phi$ is the output.
First, let us consider the simple case where $a_i$ and $h_i$ are independent, Gaussian random variables\footnote{The word ``Gaussian" happens to appear many times in this example. To clarify: though $a$ and $h$ are both Gaussian by construction, $a h$ is not, and $\phi$ is Gaussian in the CLT limit.},
\begin{equation}
    P(\vec{a},\vec{h}) = P_{\text{ind}}(\vec{a},\vec{h}) \nonumber 
    = (2 \pi \sigma_a \sigma_h)^{-N} \exp \Bigl(-\frac{1}{2\sigma_a^2} \sum_{i=1}^N a_i^2 -\frac{1}{2\sigma_h^2} \sum_{i=1}^N h_i^2 \Bigr), \\
\end{equation}
where $\sigma_a$ and $\sigma_h$ are positive and finite. Since $a_i$ and $h_i$ are independent, so are $X_i$. The CLT applies and $\phi$ is Gaussian.

Next, we will perturb $P(\vec{a},\vec{h})$ to break independence. To that end, we introduce an auxiliary random variable $H$ and define,
\begin{align}
    P(\vec{a},\vec{h}, H) &= P_{\text{ind}}(\vec{a},\vec{h},H) \nonumber \\
    &= (2 \pi \sigma_a \sigma_h)^{-N} (\sqrt{2\pi} \sigma_h)^{-1} \exp \Bigl(-\frac{1}{2\sigma_a^2} \sum_{i=1}^N a_i^2 -\frac{1}{2\sigma_h^2} \sum_{i=1}^N h_i^2 - \frac{1}{2\sigma_h^2} H^2 \Bigr),
\end{align}
where we set the standard deviation of $H$ to $\sigma_h$, for simplicity. We then define a correction term,
\begin{equation}
    P_{\text{corr}}(\vec{a},\vec{h},H) = P_{\text{ind}}(\vec{a},\vec{h},H) \cdot \exp \Bigl( -\frac{1}{2\sigma_h^2} \sum_{i=1}^N (h_i-H)^2 \Bigr).
\end{equation}
Finally, putting these together we define,
\begin{equation}
    P(\vec{a},\vec{h}, H;\alpha) = (1-\alpha) P_{\text{ind}}(\vec{a},\vec{h}) + \alpha P_{\text{corr}}(\vec{a},\vec{h}).
\end{equation}
When $\alpha=0$, the second term vanishes and both $a_i$ and $h_i$ are independent. As we turn on $\alpha>0$, the $a_i$ remain independent, but correlations are induced between the $h_i$ through a direct coupling to $H$ in $P_{\text{corr}}(\vec{a},\vec{h})$.

To quantify the non-Gaussianity of $\phi$ as a function of $\alpha$, we compute the CGF,
\begin{align}
    W_\phi[J] &= \log \mathbb{E}[e^{J \sum_i X_i/\sqrt{N}}] \nonumber \\
    &= \log \int \prod_{i=1}^{N} da_i dx_i P(\vec{a},\vec{h},H;\alpha) e^{J \sum_i a_i h_i/\sqrt{N}}.
    \label{eqn:NGexCGF}
\end{align}
As $P(\vec{a},\vec{h},H;\alpha)$ is Gaussian, \eqref{eqn:NGexCGF} can be evaluated analytically to give
\begin{equation}
    W_\phi[J] = \log \Biggl[ (1-\alpha) \Biggl(\frac{N}{N-J^2 \sigma_a^2 \sigma_h^2}\Biggr)^{N/2} + \alpha \Biggl( \frac{N^{N/2}(N-J^2 \sigma_a^2 \sigma_h^2)^{\frac{1-N}{2}}}{\sqrt{N-(N+1)J^2 \sigma_a^2 \sigma_h^2}} \Biggr) \Biggr].
\end{equation}
The odd cumulants vanish, as the $\phi$ ensemble has a $\mathbb{Z}_2$ symmetry $\phi \to -\phi$ (due to evenness of P(a)), while the even cumulants $\kappa_r^\phi$ can be computed by taking derivatives of $W_\phi[J]$. For example, the second and the fourth cumulants are
\begin{align}
    \kappa_2^\phi &= \sigma_a^2 \sigma_h^2 (1+\alpha), \\
    \kappa_4^\phi &= \sigma_a^4 \sigma_h^4 \Bigl( 9\alpha -3 \alpha^2 + \frac{6+12\alpha}{N}  \Bigr). 
\end{align}
In the limit $N \to \infty, \alpha \to 0$, the second cumulant is finite while all higher cumulants vanish, and $\phi$ is Gaussian as expected. At finite $\alpha>0$, all even cumulants are finite and in general nonzero. The ability to tune $\alpha$ thus allows one to control the degree of non-Gaussianity of $\phi$. Note that breaking independence in the large $N$ limit is not a particularly efficient way to sample from a non-Gaussian distribution of a single variable.

\subsection{Connected Correlators in NN-FT} \label{sec:NG_continuum}
We wish to establish that the ideas exemplified above --- that non-Gaussianities may arise via finite-$N$ corrections or independence breaking --- generalize to continuum NN-FT. 

In outline, one may think of this conceptually as passing from a single random variable $\phi$ ($0$d field theory) to a discrete number of random variables $\phi_i$ (lattice field theory), and finally to a continuous number of random variables $\phi(x)$ (continuum field theory), where $x\in \bR^d$. This is a textbook procedure in the context of the function-space path integral. Here we wish to instead emphasize the general procedure and parameter space perspective.

Consider the case that the continuum field $\phi(x)$ is built out of neurons $h_i(x)$ \cite{halverson2021building} as
\be 
\phi(x) = \frac{1}{\sqrt{N}} \sum_{i=1}^N h_i(x).
\ee 
If the $h_i(x)$ are independent, the CLT states that $\phi(x)$ is Gaussian in the limit $N \to \infty$. This is the essence of the NNGP correspondence.

Motivated by the single variable case, we will study non-Gaussianities arising from both finite-$N$ corrections and breaking of the independence condition. The cumulant generating functional of $\phi(x)$ is
\be \label{eqn:continuum_cgf}
W_\phi[J] = \log Z_\phi[J] = \sum_{r=1}^\infty \int \prod_{i=1}^{r} d^d x_i  \frac{J(x_1)\dots J(x_r)}{r!} \, G_{c}^{(r)}(x_1,\dots,x_r),
\ee 
where we have performed a series expansion in terms of the cumulants, a.k.a. the connected correlation functions $G_{c}^{(r)}$ of $\phi$. This is a straightforward generalization of \eqref{eqn:singlevarcgf} to the continuum. When the odd-point functions vanish the connected four-point function is 
\begin{equation}
G^{(4)}_{c}(x_1,\dots,x_4) = G^{(4)}(x_1,\dots,x_4)-(G^{(2)}(x_1,x_2)G^{(2)}(x_3,x_4) + \text{2 perms}),
\end{equation}
which will capture leading-order non-Gaussianities in many of our examples.

In the following, we will quantify non-Gaussianities in terms of non-vanishing cumulants, as well as directly in the action via an Edgeworth expansion.

\subsubsection{Finite-$N$ Corrections with Independent Neurons} \label{sec:continuumfiniteN}
We first study non-Gaussianities in the case where the neurons $h_i(x)$ are i.i.d. but $N$ is finite, e.g. single hidden layer networks, shown in \cite{hanin2023random}. We can express the CGF \eqref{eqn:continuum_cgf} in terms of the connected correlation functions of the neurons,
\begin{align} \label{eqn:independent_cgf}
W_{\phi(x)}[J] &= \log \mathbb{E}\Bigl[ \exp{\Bigl(\frac{1}{\sqrt{N}} \sum_{i=1}^{N} \int d^dx J(x) h_i(x) \Bigr) } \Bigr] \nonumber \\
&= \sum_{r=1}^{\infty} \int \prod_{i=1}^{r} d^d x_i \, \, \frac{ J(x_1)\cdots J(x_r) }{r! } \, \frac{G_{c,h_i}^{(r)}(x_1, \cdots , x_r )}{N^{r/2-1}}.
\end{align}
This result relies on the fact that for independent $h_i$, the expectation of the product is the product of the expectations, which turns the first expression into a sum on neuron CGFs. For identically distributed neurons the sum gives a factor of $N$, and the normalization $1/\sqrt{N}$ gives the $r$-dependent $N$-scaling. 
This result lets us express the connected correlators of $\phi(x)$ in terms of the connected correlators of $h_i(x)$, 
\be 
G^{(r)}_{c} (x_1, \cdots , x_r ) = \frac{G^{(r)}_{c, h_i} (x_1, \cdots , x_r ) }{N^{r/2-1}}.
\ee 
This $N$-scaling implies 
\begin{equation}
\lim_{N\to \infty}     G^{(r>2)}_{c} (x_1, \cdots , x_r ) = 0,
\label{eqn:continuum_gaussianity}
\end{equation} 
establishing Gaussianity in the limit.

\paragraph{Examples: Single Layer Cos-net and ReLU-net} We will now consider two single hidden layer architectures with finite $N$ and i.i.d. parameters. 
While the methods we describe in this section can be employed to study neural networks with arbitrary depth $L>1$, inducing statistical correlations among neurons \cite{hanin2023random}, single hidden layer architectures suffice to demonstrate their utility.

\paragraph{ReLU-net} 
First, we will consider an architecture with a single hidden layer and ReLU activation functions.
As ReLU activations are ubiquitous in machine learning applications, this is a natural example to study. Consider
\begin{equation}
\phi(x) = W^1_{i} R(W^0_{ij} x_{j} )~~\text{where}~ R(z) = \begin{cases} z,~\text{for}~ z\geq 0 \\ 0,~ \text{otherwise} \end{cases} ,
\end{equation}
with $\din = \dout = 1$, $W^{0} \sim \mathcal{N}(0, \frac{ \sigma^2_{W_0} }{\din})$, $W^{1} \sim \mathcal{N}(0, \frac{ \sigma^2_{W_1} }{ N})$.

We compute the two-point function in the parameter space description \ref{eqn:psG} to obtain
\begin{align} \label{eqn:g2conmain}
    G^{(2)}_{c,\text{ReLU}}(x,y) = \frac{\sigma_{W_0}^2\sigma_{W_1}^2}{2} \Big( R(x)R(y) + R(-x)R(-y) \Big),
\end{align}
which has a factorized structure in the terms that one might call bi-local: the function depends independently on $x$ and $y$, regardless of any relation between them.  This result is exact and does not receive $1/N$ corrections.  Non-Gaussianities induced by $1/N$ corrections manifest as a nonzero $4$-pt connected correlation function,
\begin{align}
\label{eqn:g4conmain}
G^{(4)}_{c,\text{ReLU}}(x_1, x_2, x_3, x_4) =& ~\frac{1}{N}\Bigg( \frac{15 \sigma^4_{W_0} \sigma^4_{W_1}}{4\din^2} \Big(\sum\limits_{j=\pm 1} R(j x_1)R(jx_2)R(jx_3)R(jx_4)\Big) \nonumber \\
&- \frac{\sigma^4_{W_0} \sigma^4_{W_1}}{4\din^2 }\Big( \sum\limits_{\mathcal{P}(abcd)} \sum\limits_{j= \pm 1} R(j x_a)R(j x_b)R(-j x_c)R(-j x_d) \Big) \Bigg) .
\end{align}
As expected, $G^{(4)}_{c,\text{ReLU}}(x_1, x_2, x_3, x_4)$ scales as $1/N$.

\paragraph{Cos-net} Next, let us study a single hidden layer network with cosine activation functions. The NN-FT associated to Cos-net (and its generalizations) is Euclidean invariant \cite{halverson2021building}, which is interesting on physical grounds, e.g. to satisfy one of the Osterwalder-Schrader axioms to establish an NN-QFT.
Euclidean invariance may be established using the mechanism of  \cite{maiti2021symmetryviaduality} for determining symmetries from parameter space correlators, which absorbs symmetry transformations into parameter redefinitions, yielding invariant correlators when the relevant parameter distributions are invariant under the symmetry. 

Cos-net was defined in  \cite{halverson2021building}, where its $2$-point function and connected $4$-point function were also computed. The architecture is 
\begin{equation} \label{eqn:cnetarch}
    \phi(x) = W^1_{i} \cos(W^0_{ij} x_{j} + b^0_{i})
\end{equation}
where $W^1 \sim \mathcal{N}(0, \sigma^2_{W_1} / N)$, $W^0 \sim \mathcal{N}(0, \sigma^2_{W_0} / d)$, and $b^0 \sim \text{Unif}[-\pi, \pi]$. As before, the correlation functions are computed in parameter space \eqref{eqn:psG}. The $2$-pt function 
\begin{equation}
\begin{split}
G^{(2)}_{c,\text{Cos}}(x_1, x_2) &= \frac{\sigma_{W_1}^2}{2} e^{-\frac{1}{2\din}\sigma_{W_0}^{2} (\Delta x_{12})^2}\label{eqn:cnetG2NGP}
\end{split}
\end{equation}
is manifestly translation invariant, with $\Delta x_{12} = x_1 - x_2$. The $4$-pt correlation function is
\begin{equation}
\begin{split}
  &  G^{(4)}_{c,\text{Cos}}(x_1, x_2, x_3, x_4) = \frac{\sigma_{W_1}^4 }{8  N } \sum\limits_{\mathcal{P}(abcd)}  \bigg( 3 e^{-\frac{\sigma_{W_0}^2 (\Delta x_{ab} + \Delta x_{cd} )^2}{2\din }} -2 e^{-\frac{\sigma_{W_0}^2 \left((\Delta x_{ab})^2+(\Delta x_{cd})^2\right)}{2\din }} \bigg) ,\label{eqn:cnetG4NGP}
\end{split}
\end{equation}
where $\Delta x_{ij} := x_i - x_j$ and $\mathcal{P}(abcd)$ denotes the three independent ways of drawing pairs $(x_a, x_b), (x_c, x_d)$ from the list of external vertices $(x_1, x_2, x_3, x_4)$.

We see the manifest Euclidean invariance of these correlators, and that non-Gaussianities are encoded in $G^{(4)}_{c,\text{Cos}}$ as a $1/N$ corrections.

\subsubsection{Generalized Connected Correlators from Independence Breaking} \label{sec:continuum_ib}

We now wish to generalize our theories and connected correlators to including the possibility that non-Gaussianities arise not only from $1/N$-corrections, but also from independence breaking, e.g. by developing correlations between the neurons $h_i(x)$. Previously, \cite{halverson2021building, hanin2023random} studied mixed non-Gaussianities at finite $N$ and statistical correlations among neurons.

Generalizing our approach from section \ref{sec:independence_breaking}, we parameterize breaking of statistical independence by promoting the distribution of neurons $P(h)$ to depend on a vector of hyperparameters $\vec\alpha \in \bR^q$, $P(h; \vec{\alpha})$. 

Since independence is necessary for Gaussianity via the CLT, and we will sometimes wish to perturb around the Gaussian fixed point, we require 
\be
P(h;\vec\alpha = \vec 0) = \prod_i P(h_i),
\ee 
where the hyperparameter vector $\vec \alpha$ must be chosen as part of the architecture definition.
From this expression, the neurons become independent when $\vec \alpha = 0$.

For a general $P(h;\vec{\alpha})$, the CGF is
\be
W_{\phi}[J] = \log \Bigg[ \int \Big(\prod_{i=1}^{N} D h_i \Big) P(h; \vec{\alpha}) e^{ \frac{1}{\sqrt{N}} \sum\limits_{i = 1}^{N} \int dx\, h_i(x) J(x)  } \Bigg]. \\
\ee
For small values of $\alpha$, we can expand $P(h;\vec{\alpha})$,
\be
P(h;\vec{\alpha}) = P(h;\vec{\alpha} = 0) + \sum_{r=1}^{\infty} \sum_{s_1, \cdots, s_r = 1}^{q} \frac{\alpha_{s_1} \cdots \alpha_{s_r} }{r!} \partial_{\alpha_{s_1}} \cdots \partial_{\alpha_{s_r}} P(h ; \vec{\alpha}) \Big|_{\vec{\alpha} =0 }.
\ee
Analogous to the single variable case, we define
\be
\mathcal{P}_{r, \{s_1, \cdots, s_r \}} := \frac{1}{P(h ; \vec{\alpha})} \partial_{\alpha_{s_1}} \cdots \partial_{\alpha_{s_r}} P(h ; \vec{\alpha})
\ee
satisfying the recursion relation 
\begin{align}
\mathcal{P}_{r +1, \{s_1, \cdots, s_{r +1} \}} = \frac{1}{r+1}\sum_{\gamma = 1}^{r + 1} (\mathcal{P}_{1, s_{\gamma}} + \partial_{\alpha_{s_{\gamma}}}) \mathcal{P}_{r, \{s_1, \cdots, s_{r +1} \}\backslash s_{\gamma} } .
\end{align}
Finally, we can expand \eqref{eqn:continuum_cgf} in $\vec{\alpha}$,
\be
W_{\phi}[J] = \log \Bigg[ e^{W_{\phi,\vec\alpha=0}[J] } + \sum_{r=1}^{\infty} \sum_{s_1, \cdots, s_r = 1}^{q} \frac{ \alpha_{s_1} \cdots \alpha_{s_r} }{r!}\prod_{i=1}^{N}\mathbb{E}_{P_i(h_i)} \Big[e^{ \frac{1}{\sqrt{N}}\int d^dx \, h_i(x)J(x)  } \cdot \mathcal{P}_{r, \{s_1, \cdots , s_r \} } \big|_{\vec{\alpha} = 0} \Big] \Bigg],
\label{eqn:continuum_cgf_expansion}
\ee
where $W_{\phi,\vec{\alpha} =0 }[J]$ is given in \eqref{eqn:independent_cgf}. This form of $W_{\phi}[J]$ makes it clear how one can tune $N$ and $\vec{\alpha}$ to generate and manipulate non-Gaussianities; for details see Appendix \eqref{app:CGFedgeworth}. 

For appropriately small independence breaking hyperparameter $\vec{\alpha}$, and other attributes of the architecture, the ratio of second term to first term in the logarithm of \eqref{eqn:continuum_cgf_expansion} is small. In such cases, one can approximate \eqref{eqn:continuum_cgf_expansion} using Taylor series expansion $\log(1 + x) \approx x$ around $x=0$. The CGF becomes
\begin{align}
    W_{\phi}[J] = W_{\phi,\vec\alpha=0}[J]  + \sum_{s=1}^{q} \frac{\alpha_s}{e^{W_{\phi,\vec\alpha=0}[J] }}\prod_{i=1}^{N}\mathbb{E}_{P_i(h_i)} \Big[e^{ \frac{1}{\sqrt{N}}\int d^dx \, h_i(x)J(x)  } \cdot \mathcal{P}_{1, s  } \big|_{\vec{\alpha} = 0} \Big] ,
\end{align}
and the cumulants 
\begin{align}
    G^{(r)}_{c}(x_1, \cdots , x_r) &= \frac{\partial^r W_{\phi}[J]}{\partial J(x_1) \cdots \partial J(x_r)} \Big|_{J=0} \nonumber , \\
    &= G^{(r), \text{i.i.d.}}_{c} + \vec{\alpha} \cdot \Delta G^{(r)}_{c} + O(\vec{\alpha}^2). \label{eqn:iidbreakcumulantsalpha}
\end{align}
are proportional to $\vec{\alpha}$ at the leading order. The leading order expression in $\vec{\alpha}$ is evaluated in \eqref{app:maintextderivation}.

\paragraph{Example: Single Layer Cos-net}

Let us exemplify the non-Gaussianities generated by statistical independence breaking of a single layer Cos-net architecture given in \eqref{eqn:cnetarch}.
We can break this independence by modifying the distribution from which the weights $W^0_{ij}$ (an $N\times \din$ matrix) are sampled
\begin{align}
    P(W^0) = c \exp{\Bigg[-\sum_{i,j} \Big( \frac{\din}{2 \sigma_{W_0}^2} (W^0_{ij})^2 + \frac{\alpha_{\text{IB}}}{N^2} \sum_{i_1, j_1, i_2, j_2 } (W^0_{i_1j_1})^2 (W^0_{i_2j_2})^2 \Big) \Bigg]}, 
\end{align}
where $c$ is a normalization constant. The rotational invariance preserving term $ \frac{ \aIB(\text{Tr}({W^0}^T W^0))^2}{N^2}$ introduces mixing between the weights $W^0_{ij}$ and parametric independence is explicitly broken. The degree of independence breaking can be controlled by tuning $\aIB$.  

We wish to compute the connected correlation functions to quantify the non-Gaussianities generated by independence breaking. In general, this is a difficult problem. However, when $\aIB \ll 1$, we can perform a perturbative expansion in $\aIB$. Setting $\din=1$ for simplicity, we obtain
\begin{align} 
    G^{(2)}_{c,\text{Cos}} (x_1, x_2) =& \frac{\aIB  \sigma_{W_0}^4 \sigma_{W_1}^2 e^{-\frac{\sigma_{W_0}^2 (\Delta x_{12})^2}{2 }} }{2 N}\bigg[ - \frac{\left(1 -5  \sigma_{W_0}^2 (\Delta x_{12})^2+\sigma_{W_0}^4 (\Delta x_{12})^4\right)}{ N}   \nonumber \\
    &  +\sigma_{W_0}^2 (\Delta x_{12})^2 \bigg], \label{eqn:cnet1} 
\end{align}
\begin{align}
    &G^{(4)}_{c,\text{Cos}}(x_1, \cdots, x_4) = G^{(4)}_{\substack{c, \text{Cos} \\ \text{i.i.d.}}} (x_1, \cdots, x_4) + \frac{\aIB \sigma_{W^0}^4 \sigma_{W^1}^4}{8  N^2} \sum\limits_{\mathcal{P}(abcd)} \Bigg[ 6 - \Big(2 \sigma_{W^0}^2 (\Delta x_{ab}^2 +\Delta x_{cd}^2)  \nonumber \\
    &  + 2 \sigma_{W^0}^4 \Delta x_{ab}^2\Delta x_{cd}^2 \Big) e^{-\frac{\sigma_{W_0}^2 }{2 }\left(\Delta x_{ab}^2+\Delta x_{cd}^2\right)} + \Big(3 + 3 \sigma_{W^0}^2 (\Delta x_{ab} + \Delta x_{cd} )^2 \Big)e^{-\frac{\sigma_{W_0}^2 }{2}(\Delta x_{ab} + \Delta x_{cd} )^2} \Bigg], \label{eqn:cnet2}
\end{align}
to leading order in $\aIB$, where $G^{(4)}_{\substack{c, \text{Cos} \\ \text{i.i.d.}}} (x_1, \cdots, x_4)$ is obtained at $\din =1$ from \eqref{eqn:cnetG4NGP}.
Non-Gaussianities at finite $N$, and $\aIB \neq 0$ still preserve the translation invariance of the $2$nd and $4$th cumulants of Cos-net architecture.  

We refer the reader to Appendix \eqref{appendix:cosnet_eg} for details, where we also compute leading order non-Gaussian corrections to first two cumulants in a single hidden layer Gauss-net at $\aIB \neq 0$, finite $N$, for $\din=1$.

\section{Computing Actions from Connected Correlators \label{sec:actions}}
In Section \ref{sec:NG} we systematically studied non-Gaussianities in neural network field theories by parametrically violating two assumptions of the Central Limit Theorem: infinite-$N$ and independence. The study was performed at the level of connected correlators, rather than actions, because every NN-FT admits a parameter space description of connected correlators, even if an action is not known.

In this section we will develop these techniques for calculating actions from connected correlators, including in terms of Feynman diagrams in which the connected correlators are vertices. More specifically:
\begin{itemize}
\item \textbf{Field Density from Connected Correlators: Edgeworth Expansion.} In Section \ref{sec:Edgeworth} we review how knowledge of the cumulants of a single random variable may be used to approximate its probability density, and then we generalize to the field theory case, which has a continuum of random variables. This gives an expression for $P[\phi]=\exp(-S[\phi])$ in terms of connected correlation functions. We present an explicit example in the case of a single variable.
\item \textbf{Computing the Action with Feynman Diagrams.}  Given the Edgeworth expansion, we develop a method to compute the action perturbatively via Feynman diagrams, which becomes clear due to a formal similarity between the Edgeworth expansion and the partition function of a field theory. This is a result that is applicable to general field theories.
\item \textbf{Neural Network Field Theory Actions.} In Section \ref{sec:edgeworth-nnft} we specify the analysis of Section \ref{sec:ContinuumEdgeworth} to the case of neural network field theories. We derive the leading order form of the action for the case of non-Gaussianities induced either by $1/N$-corrections or independence breaking.
\item \textbf{Neural Network Field Theory Examples.}
In Section \ref{sec:Feynmanexamples} we derive the leading-order action in $1/N$ for concrete neural network architectures.
\end{itemize}

\subsection{Field Density from Connected Correlators: Edgeworth Expansion \label{sec:Edgeworth}}
The Edgeworth expansion from statistics (see e.g. \cite{GVK025155202} for a textbook statistics description and \cite{Naveh_2021} for an ML study) can be used to construct the probability density from the cumulants.
The key observation which allows the Edgeworth expansion to be applied in a
field theory is that the normal relation for the generating function in terms
of the action
\begin{equation}
  e^{ W [J]} = \int d \phi \, e^{- S [\phi] + J \phi}
\end{equation}
can be inverted to express the action in terms of the generating functional.
Adding a source term in the exponent, mapping $J\to i J$ and integrating over $J$, we have
\begin{equation}
  \int d J e^{ W [i J] - i J \phi} = \int d J e^{ - i J \phi} \int d
  \phi' e^{- S [\phi'] + i J \phi'} =  e^{- S [\phi]}
\end{equation}
where
\begin{equation}
  \int d J e^{i J (\phi' - \phi)} = \delta [\phi' - \phi]
\end{equation}
has been used. Deforming the $J$ integration contour back to real $J$ then results in
\begin{equation}
  P [\phi] = e^{- S [\phi]} = \int d J e^{ W [J] - J \phi},
\end{equation}
This gives the probability density and action in terms of $W [J]$. This result can also be thought of as arising from an inverse Fourier transform of the characteristic function.

Then to apply the Edgeworth expansion for a single random variable $\phi$, we write $W [J]$ in terms of cumulants
\begin{equation}
  W [J] = \sum_{r = 1}^{\infty} \frac{\kappa_r}{r!} J^r,
\end{equation}
which lets us write
\begin{align}
  P[\phi] &=  \exp \left[ \sum_{r = 3}^{\infty}
  \frac{\kappa_r}{r!} (-\partial_{\phi})^r \right] \int d J e^{ \kappa_2
  \frac{J^2}{2} + \kappa_1 J - J \phi}, \nonumber \\
&  =  \exp \left[ \sum_{r = 3}^{\infty} \frac{\kappa_r}{r!} (-
  \partial_{\phi})^{r} \right] e^{ -\frac{(\phi - \kappa_1)^2}{2 \kappa_2}},
\end{align}
where the Gaussian integral has been performed by mapping $J\to iJ$ (alternatively, working with the characteristic function the whole time) and we have neglected the normalization factor.
We have an expression for the density $P_\phi$ as an expansion around the Gaussian with mean $\kappa_1$ and variance $\kappa_2$.

The result may be extended to the field theory case, where $\phi$ is replaced by $\phi(x)$, a continuum of mean free random variables. Then the relation is
\begin{equation}\label{eqn:continuum_edgeworth_2}
e^{-S[\phi]} = \frac{1}{Z} \exp{ \Big(\sum_{r=3}^{\infty} \frac{(- 1)^r}{r!} \int \prod_{i=1}^{r} d^d x_i G_{{c}}^{(r)} (x_1, \cdots , x_r ) \frac{\delta}{\delta \phi(x_{1}) } \cdots \frac{\delta}{\delta \phi(x_{r})}\Big)} e^{-S_G[\phi]},
\end{equation}
where the Gaussian Process action $S_G$ is defined as
\begin{equation}
    S_G[\phi] = \frac{1}{2} \int d^dx_1 d^dx_2\, \phi(x_1) G_{{c}}^{(2)}(x_1, x_2)^{-1}  \phi(x_2),
\end{equation}

To the extent that there is a perturbative ordering to the correlators
through
some expansion parameter (such as $\frac{1}{N}$ or independence breaking),
this expression can be evaluated perturbatively to systematically construct an
action from the cumulants.
\footnote{
In the finite-dimensional version of the Edgeworth expansion, it is sometimes
convenient to further express the powers of derivatives in terms of
probabilist's Hermite polynomials using
\begin{equation}
(-\partial_x)^r\, e^{-\frac{x^2}{2}}=: H_r(x)\, e^{-\frac{x^2}{2}},
\end{equation}
In the field theory case, using Hermite polynomials provides no obvious advantage. }

\subsubsection*{1D Example: Sum of $N$ Uniform Random Variables} \label{sec:SingleVarEdgeworthExample}
Let us demonstrate the Edgeworth expansion in a simple example. Consider the standardized sum of $N$ i.i.d. random variables sampled from a uniform distribution
\be
\phi = \frac{1}{\sqrt{N}} \sum_{i=1}^N X_i, \quad X_i \sim \text{Unif}(-1/2, 1/2) \, \, \forall i.
\ee
The cumulants of $X_i$ are
\begin{align}
 \kappa_1^{X_i} &= 0, \\
 \kappa_r^{X_i} &= \frac{B_r}{r} \, \, \text{for} \, \, r \geq 2,
\end{align}
where $B_r$ is the $r^\text{th}$ Bernoulli number.\footnote{$B_r$ vanishes for odd $r\geq2$. First few nonzero Bernoulli numbers are: $B_2=\frac{1}{6}, B_4=-\frac{1}{30}, B_6=\frac{1}{42}$.} Plugging this into \eqref{eqn:phicumulants_N_scaling}, the cumulants of $\phi$ are
\begin{equation} \label{eqn:singlevarcumulants}
 \kappa_r^{\phi} = \frac{B_r}{r N^{r/2-1}}.
\end{equation}
At finite $N$, the cumulants $\kappa_{r>2}^\phi$ are nonzero and $\phi$ is non-Gaussian. Using these cumulants, we can write down the probability distribution function of $\phi$ via an Edgeworth expansion,
\begin{align}
    P_\phi &= \frac{1}{Z} \exp{\Biggl[\sum_{r=3}^\infty \frac{\kappa_r^\phi}{r!} (-\partial_\phi)^r}\Biggr] e^{- \phi^2/2\kappa_2^\phi} \nonumber \\
    &= \frac{1}{Z} \exp{\Biggl[\sum_{r=3}^\infty \frac{B_r}{r! r N^{r/2-1}} (-\partial_\phi)^r}\Biggr] e^{- \phi^2/B_2}
\end{align}
Truncating the sum at $r=4$, expanding the exponential, and keeping terms up to $O(1/N)$ we get
\begin{align} \label{eqn:singlevarexampleEdgeworth}
P_\phi &= \frac{1}{Z} \Bigg[1 + \kappa_4^\phi \Big(\frac{1}{8 (\kappa_2^\phi)^2} - \frac{1}{4 (\kappa_2^\phi)^3} \phi^2 + \frac{1}{24 (\kappa_2^\phi)^4} \phi^4 \big) + O(1/N^{3/2}) \Bigg] e^{- \phi^2/2\kappa_2^\phi}, \nonumber \\
&= \frac{1}{Z'} \exp \Bigg[ -\Big(\frac{1}{2\kappa_2^\phi} + \frac{\kappa_4^\phi}{4 (\kappa_2^\phi)^3}\Big)\phi^2 + \frac{\kappa_4^\phi}{24 (\kappa_2^\phi)^4} \phi^4 + O(1/N^{3/2}) \Bigg], \nonumber \\
&= \frac{1}{Z'} \exp \Bigg[ \Big(-6 + \frac{18}{5N}\Big)\phi^2 - \frac{36}{5N} \phi^4 + O(1/N^{3/2}) \Bigg],
\end{align}
where on the second line we absorbed the constant term into the normalization constant $Z'$. At order $O(N^0)$, the exponent in \eqref{eqn:singlevarexampleEdgeworth} is quadratic and $\phi$ is Gaussian distributed. Gaussianity is then broken by a quartic interaction at order $O(1/N)$.

It is worth noting that the cumulants of $\phi$ are given by simple closed form expressions, see Equation \eqref{eqn:singlevarcumulants}, while $P_\phi$ involves a perturbative expansion in $1/N$. This is in contrast to weakly coupled field theories, where we often start from a simple action expressed in closed form and calculate the connected correlation functions via a perturbative expansion in the coefficients of interaction terms.

\subsection{Computing the Action with Feynman Diagrams
\label{sec:ContinuumEdgeworth}}
In a field theory a powerful tool for organizing a perturbation expansion is with Feynman diagrams. Just as Feynman diagrams can be used to compute the cumulants perturbatively in an expansion parameter from an action, they can also be used to compute the action perturbatively from the cumulants.
To understand the derivation, recall the expression 
for the partition function
\begin{align}\label{eqn:continuum_partition_function}
    e^{W[J]} &= Z[J] = c' \exp{ \Big(\sum_{r=3}^{\infty} \int \prod_{i=1}^{r} d^d x_i \, g_{r} (x_1, \cdots , x_r ) \frac{\delta}{\delta J(x_{1}) } \cdots \frac{\delta}{\delta J(x_{r})}\Big)} e^{-S_{0}[J]},
\end{align}
where we have introduced couplings $g_r$ instead of $g_r/r!$,
\begin{equation}
    S_0[J] = \int dx_1 dx_2 J(x_1) \Delta(x_1, x_2) J(x_2),
\end{equation}
and $\Delta(x_1, x_2)$ is the free propagator. The expression \eqref{eqn:continuum_partition_function} arises 
by taking the usual expression for the partition function 
\begin{equation}
\label{eqn:usual_partition_function}
Z[J] = \int D\phi \,e^{-S_\text{free}[\phi] - S_\text{int} + \int d^d x J(x) \phi(x)}
\end{equation}
and replacing the $\phi$'s in the interaction terms
\begin{equation}
S_{\text{int}} = \sum_{r=3}^\infty  \int \prod_{i=1}^r d^dx_i \,  g_r(x_1,\dots,x_r) \, \phi(x_1)\dots \phi(x_r)
\end{equation}
by $\delta/\delta J$'s. Pulling the $J$-derivatives outside of the $\int D\phi$ in \eqref{eqn:usual_partition_function} and performing the Gaussian integral yields \eqref{eqn:continuum_partition_function}. These manipulations closely mirror the Edgeworth expansion.

\begin{table}[t]
\centering
\begin{tabular}{|c|c|c|}
\hline
& Field Picture & Source Picture \\ \hline 
Field & $\phi(x)$ & $J(x)$ \\
CGF & $W[J] = \log(Z[J])$ & $S[\phi] = -\log(P[\phi])$\\ 
Cumulant & $G^{(r)}_c(x_1,\dots,x_r)$ & $g_r(x_1,\dots, x_r)$ \\ \hline
\end{tabular}
\caption{The Edgeworth expansion for $P[\phi]$ and the interaction expansion of $Z[J]$ are formally related by a change of variables, given here up to constant factors. Due to this relationship, non-local couplings and connected correlators may both be computed by appropriate connected Feynman diagrams.}
\label{tab:field-source-duality}
\end{table}

The Edgeworth expansion \eqref{eqn:continuum_edgeworth_2} is related to the partition function \eqref{eqn:continuum_partition_function} by a simple change of variables, given in Table \ref{tab:field-source-duality}, which one might think of as a duality map between a field picture and a source picture. This relationship between the Edgeworth expansion and the partition function immediately tells us that the analog of $g_r(x_1,\dots, x_n)$ are the connected correlation functions $G_{{c}}^{(r)} (x_1, \cdots , x_r )$ in \eqref{eqn:continuum_edgeworth_2}.

We may therefore compute the couplings $g_r(x_1,\dots, x_n)$ in the same way that we compute the connected correlators $G_{{c}}^{(r)} (x_1, \cdots , x_r ).$
In a weakly coupled field theory, one can compute the connected correlation functions $G_{{c}}^{(r)} (x_1, \cdots , x_r )$ in terms of the couplings $g_r(x_1, \cdots, x_r)$ perturbatively via Feynman diagrams. An Edgeworth expansion allows us to do the converse and compute the couplings $g_r(x_1, \cdots, x_r)$ in terms of the connected correlation functions $G_{{c}}^{(r)} (x_1, \cdots , x_r )$. The similarity between \eqref{eqn:continuum_edgeworth_2} and \eqref{eqn:continuum_partition_function} suggests that the terms in the expansion for $g_r(x_1, \cdots, x_r)$ can be represented by Feynman diagrams, whose vertices are connected correlators, e.g. 
\begin{equation}
\begin{tikzpicture}
\begin{feynman}
  every vertex/.style={red, dot};
every particle/.style={blue};
    \vertex[empty dot] (m) at (0,0) {\contour{black}{}{$G^{(6)}_{c}$}};
    \vertex (a) at (-1,0.5) ;
    \vertex (b) at ( -1,-0.5);
    \vertex (c) at (1, 0.5);
    \vertex (b1) at (0,1);
    \vertex (d) at (1,-0.5);
    \vertex (a1) at (0,-1) ;
    \vertex (an) at (-0.31,0.15) ;
    \vertex (bn) at ( -0.10,-0.47);
    \vertex (cn) at (1.0, -0.15);
    \vertex (b1n) at (0.85,0.45);
    \vertex (dn) at (0.11,0.63);
    \vertex (a1n) at (0.55,-0.65) ;
     \diagram* {
      (a) -- [dashed] (m) -- [dashed](b),
      (c) -- [dashed] (m) -- [dashed] (d),
      (a1) -- [dashed] (m) -- [dashed] (b1)};
  \end{feynman}
\end{tikzpicture}
\end{equation}
in the case of a six-point vertex. Notably, the vertex is itself a function and lines enter the $n$-point vertex at $n$ locations.

To compute the coupling $g_r(x_1,\dots,x_r)$ in terms of Feynman diagrams, one sums over all connected $r$-point Feynman diagrams made out of $G^{(n)}_{c}$ vertices. By convention, we do not label internal points on the vertices, in order to simplify the combinatorics. For instance, the four-point coupling $g_4(x_1,\dots,x_4)$ has a diagram 
\begin{equation}
\begin{tikzpicture}
\begin{feynman}
  every vertex/.style={red, dot};
every particle/.style={blue};
    \vertex[empty dot] (m) at (0,0) {\contour{black}{}{$G^{(4)}_{c}$}};
    \vertex[label=left:$x_1$] (a) at (-1,1) ;
    \vertex[label=left:$x_2$] (b) at ( -1,-1);
    \vertex[label=right:$x_3$] (c) at (1, 1);
    \vertex[label=right:$x_4$] (d) at (1,-1);
    \vertex (a1) at (-0.31,0.31) ;
    \vertex (b1) at ( -0.31,-0.31);
    \vertex (c1) at (0.31, 0.31);
    \vertex (d1) at (0.31,-0.31);
    \diagram* {
      (a) -- [dashed] (m) -- [dashed](b),
      (c) -- [dashed] (m) -- [dashed] (d) };
  \end{feynman}
\end{tikzpicture}\end{equation}
where it is to be understood that connections to internal points in a vertex appear in all possible combinations. Analytic expressions may be obtained from the diagrams via the Feynman rules given in Table \ref{tab:feynrules}.
If $G_{{c}}^{(2)}(x_i, y_j)^{-1}= \frac{\delta^2 S_G}{\delta \phi(x_i) \delta \phi(y_j)}$ involves differential operators, it can be evaluated by Fourier transformation, see Appendix \eqref{app:G2inverseFT}

\begin{table}[thb]
\textbf{Feynman Rules for $g_r(x_1,\dots,x_r)$.}
\begin{enumerate}
    \item Internal points associated to vertices are unlabeled, for diagrammatic simplicity. Propagators therefore connect to internal points in all possible ways.
    \item For each propagator between $z_i$ and $z_j$, 
    \begin{align}
        \vcenter{\hbox{\begin{tikzpicture}
\begin{feynman}
  every vertex/.style={red, dot};
every particle/.style={blue};
  \vertex[label=left:$z_i$] (a1) at (-0.4,0) ;
    \vertex[label=right:$z_j$] (b1) at ( 0.4,0);
    \diagram* {
      (a1) -- [dashed] (b1)};
  \end{feynman}
\end{tikzpicture}}} = G_{{c}}^{(2)}(z_i, z_j)^{-1}.
    \end{align}
    \item For each vertex,
    \begin{align}
        \vcenter{\hbox{\begin{tikzpicture}
\begin{feynman}
  every vertex/.style={red, dot};
every particle/.style={blue};
    \vertex[empty dot] (m) at (0,0) {\contour{black}{}{$G^{(n)}_{c}$}};
    \vertex (a) at (-1,0.5) ;
    \vertex (b) at ( -1,-0.5);
    \vertex (c) at (1, 0.5);
    \vertex (b1) at (0,1);
    \vertex (d) at (1,-0.5);
    \vertex (a1) at (0,-1) ;
    \vertex (an) at (-0.31,0.15) ;
    \vertex (bn) at ( -0.10,-0.47);
    \vertex (cn) at (1.0, -0.15);
    \vertex (b1n) at (0.85,0.45);
    \vertex (dn) at (0.11,0.63);
    \vertex (a1n) at (0.55,-0.65) ;
     \diagram* {
      (a) -- [dashed] (m) -- [dashed](b),
      (c) -- [dashed] (m) -- [dashed] (d),
      (a1) -- [dashed] (m) -- [dashed] (b1)};
  \end{feynman}
\end{tikzpicture}}} = (-1)^{n}\int d^dy_1 \cdots d^dy_n \, G_{{c}}^{(n)} (y_1, \cdots , y_n ).
    \end{align}
    \item Divide by symmetry factor and insert overall $(-)$.
\end{enumerate}
\caption{Feynman rules for computing $g_r$ from each connected diagram with $G^{(n)}_c$ vertices. }
\label{tab:feynrules}
\end{table}

As an example, let us compute a contribution to the quartic coupling $g_4(x_1, x_2, x_3, x_4)$ from a $G^{(4)}_{c}$ vertex 
\begin{align}
g_4(x_1,\dots,x_4) =& \,\,
     -\frac{1}{4!} \Big[ \int dy_1 dy_2 dy_3 dy_4 \,G^{(4)}_{c} (y_1, y_2, y_3, y_4) \, G_{{c}}^{(2)}(y_1, x_1)^{-1} G_{{c}}^{(2)}(y_2, x_2)^{-1}   \nonumber \\
    & \quad G_{{c}}^{(2)}(y_3, x_3)^{-1} G_{{c}}^{(2)}(y_4, x_4)^{-1}  + \text{perms}  \Big] + \dots  \\
 = &  \qquad   \vcenter{\hbox{\begin{tikzpicture}
\begin{feynman}
  every vertex/.style={red, dot};
every particle/.style={blue};
    \vertex[empty dot] (m) at (0,0) {\contour{black}{}{$G^{(4)}_{c}$}};
    \vertex[label=left:$x_1$] (a) at (-1,1) ;
    \vertex[label=left:$x_2$] (b) at ( -1,-1);
    \vertex[label=right:$x_3$] (c) at (1, 1);
    \vertex[label=right:$x_4$] (d) at (1,-1);
    \vertex (a1) at (-0.31,0.31) ;
    \vertex (b1) at ( -0.31,-0.31);
    \vertex (c1) at (0.31, 0.31);
    \vertex (d1) at (0.31,-0.31);
    \diagram* {
      (a) -- [dashed] (m) -- [dashed](b),
      (c) -- [dashed] (m) -- [dashed] (d) };
  \end{feynman}
\end{tikzpicture}}}  + \dots, \label{eqn:EdgeworthFeynman1}
\end{align}
where the dots represent contributions from other diagrams, and ``perms'' represents other diagrams from permutations over internal points.
A combinatoric factor of $4!$ from summing over internal points cancels out the prefactor $1/4!$ from Edgeworth expansion.

The Edgeworth expansion \eqref{eqn:continuum_edgeworth_2} involves an infinite sum. Correspondingly, computing $g_r(x_1, \cdots, x_r)$ requires summing over infinitely many Feynman diagrams. When all but finitely many terms in the expansion are parametrically suppressed, the expansion can be truncated at finite order to provide an approximation of $g_r(x_1, \cdots, x_r)$. We will apply these rules to concrete examples later in this section and demonstrate how approximations to $g_r(x_1, \cdots, x_r)$ can be obtained systematically.

While our focus is on neural network field theories, we emphasize that Edgeworth expansions can be utilized in any field theory where the connected correlation functions are known, and the expansion in \eqref{eqn:continuum_edgeworth_2} is not divergent. 

\subsubsection{Example: Non-local $\phi^4$ Theory.}

Aside from any application in NN-FT, it is interesting to study the self-consistency of the Edgeworth expansion. We do so in a famous case, $\phi^4$ theory, generalized to the case of non-local quartic interactions, in order to demonstrate the ability of the Edgeworth method to handle non-locality.
Consider the action 
\begin{align} \label{eqn:phi4action}
S[{\phi}] &= \int d^dx_1 d^d x_2 \, \frac{1}{2} \phi(x_1) G_{G, \phi}^{(2)}(x_1, x_2)^{-1} \phi(x_2) \nonumber \\
&+ \frac{1}{4!} \int d^dx_1 d^dx_2 d^dx_3 d^dx_4 \lambda(x_1, x_2, x_3, x_4) \phi(x_1) \phi(x_2) \phi(x_3) \phi(x_4),
\end{align}
where $G_{G, \phi}^{(2)}(x_1, x_2)^{-1}$ and $\lambda(x_1, x_2, x_3, x_4)$ are both totally symmetric, and $G_{G, \phi}^{(2)}(x_1, x_2)^{-1}$ is the operator in the free action $S_G[\phi]$. We denote the free propagator $D(x_1, x_2)$ so that
$\int d^dx' \, G_{G, \phi}^{(2)}(x_1, x_2)^{-1} \, D(x', x_2) = \delta^d(x_1-x_2)$. We can then expand $G_{c}^{(2)}(x_1, x_2)$ in $\lambda(x_1, x_2, x_3, x_4)$, and at leading order,
\begin{align} \label{eqn:phi4G2con}
    G_{c}^{(2)}(x_1, x_2) = D(x_1, x_2) + \frac{1}{2} \int d^dy_1 \cdots d^dy_4 \lambda(y_1, y_2, y_3, y_4) D(x_1, y_1)D(y_2, y_3) D(y_4, x_2) ,
\end{align}
where the $\frac{1}{2}$ is a symmetry factor. Similarly,
\begin{align} \label{eqn:phi4G4con}
    G^{(4)}_{c}(x_1, \cdots, x_4) =&\, \int d^dx'_1 \cdots d^dx'_4 \, \lambda(x'_1, x'_2, x'_3, x'_4) D(x_1, x'_1) D(x_2, x'_2) D(x_3, x'_3) D(x_4, x'_4) \nonumber \\
    &+ O(\lambda^2).
\end{align}
There are no other connected correlators that have contributions at $O(\lambda)$. To perform an Edgeworth expansion, we first need to write down the inverse propagator,
\be \label{eqn:phi4inverseprop}
G_{{c}}^{(2)}(x_1, x_2)^{-1} = G_{G, \phi}^{(2)}(x_1, x_2)^{-1} - \frac{1}{2} \int d^dx_3 d^dx_4 \lambda(x_1, x_2, x_3, x_4) D(x_3, x_4) + O(\lambda^2).
\ee
Given \eqref{eqn:phi4inverseprop}, it is easy to verify that
\be
\int dx' G_{{c}}^{(2)}(x_1, x')^{-1} G^{(2)}_{{c}} (x', x_2) = \delta(x_1 - x_2) + O(\lambda^2).
\ee
At this point, let us introduce a shorthand notation to improve readability, rewriting $\int d^dx_1 d^dx_2\,G_{c}^{(2)}(x_1, x_2)$, $\int d^dx_1 d^dx_2\,G_{{c}}^{(2)}(x_1, x_2)^{-1}$ and $\int d^dx_1 \cdots d^dx_4\,G^{(4)}_{c}(x_1, \cdots, x_4)$ as,
\begin{align}
    G_{xy} &= D_{xy} + \frac{1}{2} \lambda_{1234}D_{1x}D_{23}D_{4y} + O(\lambda^2) , \\
    G_{xy}^{-1} &= G_{G, \phi}^{(2)}(x, y)^{-1} - \frac{1}{2} \lambda_{xy12}D_{12} + O(\lambda^2), \\
    G_{1234} &= \lambda_{1'2'3'4'}D_{1'1}D_{2'2} D_{3'3} D_{4'4} + O(\lambda^2),
\end{align}
respectively. Finally, we obtain the Edgeworth expansion at $O(\lambda)$ by plugging in \eqref{eqn:phi4inverseprop} and \eqref{eqn:phi4G4con} into \eqref{eqn:continuum_edgeworth_2},
\be
P[\phi] = \frac{1}{Z} \exp{\Bigl( \frac{1}{4!}  G_{1234} \delta_1 \delta_2 \delta_3 \delta_4  \Bigr)} \exp{\Bigl( -\frac{1}{2} \phi_x G^{-1}_{xy} \phi_y \Bigr)} + O(\lambda^2),
\ee
where $\delta_1 := \delta / \delta \phi(x_1)$. Expanding the first exponential and performing the derivatives we obtain
\be
P[\phi] = \frac{1}{Z} \Big[1 - \frac{\lambda_{1234}}{8}D_{12}D_{34} - \frac{\lambda_{1234}}{4!}\phi_1 \phi_2 \phi_3 \phi_4 \Big] \exp{\Big(  - \frac{1}{2}\phi_x G_{G, \phi}^{(2)}(x, y)^{-1} \phi_y \Big)} + O(\lambda^2),
\ee
with $\lambda_{1234} := \int d^dx_1 \cdots d^dx_4 \, \lambda(x_1,\cdots, x_4)$, and $\phi_x := \phi(x)$. The second term does not depend on $\phi$ and can be absorbed into the normalization factor, resulting in 
\be
P[\phi] = \frac{1}{Z'} \exp{ \Big( - \frac{1}{2}\phi_x G_{G, \phi}^{(2)}(x_1, x_2)^{-1}\phi_y - \frac{\lambda_{1234}}{4!}\phi_1 \phi_2 \phi_3 \phi_4 \Big)}  + O(\lambda^2).
\ee
We have recovered the $\phi^4$ action at $O(\lambda)$, as expected.

\subsection{General Interacting Actions in NN-FT\label{sec:edgeworth-nnft}}

We now study the Edgeworth expansion in neural network field theories. We will modify the general analysis of the previous section to the case where non-Gaussianities are generated by the two mechanisms we described in Section \ref{sec:NG}, namely, by violating assumptions of the CLT by finite $N$ corrections and independence breaking.

\bigskip
\noindent \textbf{Interactions from $1/N$-corrections.} 
As we discussed in section \ref{sec:continuumfiniteN}, non-Gaussianities arising due to $1/N$ corrections result in connected correlation functions that scale as
\begin{equation}
    G_{{c}}^{(r)} (x_1, \cdots , x_r) \propto \frac{1}{N^{r/2-1}},
\end{equation}
for a single hidden layer network. At large $N$, the action can be approximated systematically by organizing the Edgeworth expansion in powers of $1/N$, calculating the couplings via Feynman diagrams, and truncating at a fixed order in $1/N$.  

To do so, we need to know how the couplings scale with $N$. We have studied a case in \eqref{eqn:EdgeworthFeynman1} where only the even-point correlators are non-zero, and clearly there is a $1/N$ contribution to $g_4$ from a single $G^{(4)}_{c}$ vertex; any higher order correlator $G^{(r>4)}_{c}$ contributes at $1/N^{r/2-1}$ and higher. Consider now contributions to the couplings $g_{r>4}$. There is a tree-level $1/N^{r/2-1}$ contribution from a single $G^{(r)}_{c}$ vertex and there are $1/N^{n/2-1}$ contributions from a $G^{(n>r)}_{c}$ vertex with an appropriate number of loops; both are more suppressed than the $1/N$ contribution to $g_4$. Finally, consider contributions from $V$ number of $G^{(n<r)}_{c}$ vertices. Forming a connected diagram requires $nV > r$, which implies $V\geq 2$ and therefore the contribution is of order $1/N^{\geq n-1}$, which is more suppressed than $1/N$ since $n$ begins at $3$ in the Edgeworth expansion. Therefore, the single-vertex tree-level contribution to $g_4$ is the leading contribution in $1/N$.

The quartic coupling $g_4(x_1, x_2, x_3, x_4)$, at leading order in $G_{{c}}^{(4)} \propto 1/N$, is
\begin{align}
g_4(x_1,\dots,x_4) =&
    \,\, - \frac{1}{4!} \Bigg[ \int dy_1 dy_2 dy_3 dy_4 \, G^{(4)}_{c} (y_1, y_2, y_3, y_4) \,G_{{c}}^{(2)}(y_1, x_1)^{-1} G_{{c}}^{(2)}(y_2, x_2)^{-1}   \nonumber \\
    & \quad G_{{c}}^{(2)}(y_3, x_3)^{-1} G_{{c}}^{(2)}(y_4, x_4)^{-1} \, + \text{perms} \Bigg] + O\left(\frac{1}{N^2}\right), \\
 = & \quad  \vcenter{\hbox{\begin{tikzpicture}
\begin{feynman}
  every vertex/.style={red, dot};
every particle/.style={blue};
    \vertex[empty dot] (m) at (0,0) {\contour{black}{}{$G^{(4)}_{c}$}};
    \vertex[label=left:$x_1$] (a) at (-1,1) ;
    \vertex[label=left:$x_2$] (b) at ( -1,-1);
    \vertex[label=right:$x_3$] (c) at (1, 1);
    \vertex[label=right:$x_4$] (d) at (1,-1);
    \vertex (a1) at (-0.31,0.31) ;
    \vertex (b1) at ( -0.31,-0.31);
    \vertex (c1) at (0.31, 0.31);
    \vertex (d1) at (0.31,-0.31);
    \diagram* {
      (a) -- [dashed] (m) -- [dashed](b),
      (c) -- [dashed] (m) -- [dashed] (d) };
  \end{feynman}
\end{tikzpicture}}} + O\left(\frac{1}{N^2}\right). \label{eqn:feynmanedgeworth2}
\end{align}
We may compute this coupling in a NN-FT by first computing $G^{(4)}_{c}$ in parameter space.

In summary, the leading-order in $1/N$ action for a single layer NN-FT is 
\begin{equation}
S = S_G + \int d^dx_1\dots d^dx_4\,\,\, g_4(x_1,\dots,x_4) \, \phi(x_1)\dots\phi(x_4) +   O\left(\frac{1}{N^2}\right ),
\end{equation}
where $g_4$ at $O(1/N)$ is given in \eqref{eqn:feynmanedgeworth2}, under the assumption that the odd-point functions are zero, as in the architectures of Section \eqref{sec:Feynmanexamples}.

\bigskip
\noindent \textbf{Interactions from Independence Breaking.}
Non-Gaussianities generated via independence breaking alone are qualitatively different than those from $1/N$ corrections. 

We wish to determine the leading-order action due to independence breaking.
Focusing on the case where independence breaking is controlled by a single parameter $\alpha$ for simplicity, it follows from \eqref{eqn:iidbreakcumulantsalpha}, that the connected correlation functions scale as
\begin{equation}
    G_{{c}}^{(r)} (x_1, \cdots , x_r) \propto \alpha \quad \forall r>2
\end{equation}
at $N \rightarrow \infty$ limit, since the connected correlators $G_{{c}}^{(r), \text{free}}|_{r>2}$ of the free theory vanish.

As a result, each coupling $g_r(x_1, \cdots, x_r)$ receives contributions from tree-level diagrams of all connected correlators, at leading order in $\alpha$. More generally, at any given order in $\alpha$, there are infinitely many diagrams from all connected correlators to $g_r(x_1, \cdots, x_r)$. For example, the expansion for  $g_4(x_1, x_2, x_3, x_4)$ at $O(\alpha)$ includes terms proportional to $G_{{c}}^{(2n)} (x_1, \cdots , x_{2n})$ for all $n>1$,
\begin{align}
    &g_4(x_1, x_2, x_3, x_4) = -\sum_{n=2}^{\infty} \frac{(-1)^{2n}}{(2n)!} \Big[ \int dy_1 \cdots dy_{2n} \, G_{{c}}^{(2n)} (y_1, \cdots , y_{2n}) \, G_{{c}}^{(2)}(y_1, x_1)^{-1}\,  \nonumber \\
    &\quad G_{{c}}^{(2)}(y_2, x_2)^{-1} \, G_{{c}}^{(2)}(y_3, x_3)^{-1} \, G_{{c}}^{(2)}(y_4, x_4)^{-1}\, \prod_{m=5}^{2n-1} G_{{c}}^{(2)}(y_m, y_{m+1})^{-1} + \text{perms} \Big] + O(\alpha^2),\\
= &  (-1)^{2n} \Bigg(  \vcenter{\hbox{\begin{tikzpicture}
\begin{feynman}
  every vertex/.style={red, dot};
every particle/.style={blue};
    \vertex[empty dot] (m) at (0,0) {\contour{black}{}{$G^{(4)}_{c}$}};
    \vertex[label=left:$x_1$] (a) at (-0.7,0.7) ;
    \vertex[label=left:$x_2$] (b) at ( -0.7,-0.7);
    \vertex[label=right:$x_3$] (c) at (0.7, 0.7);
    \vertex[label=right:$x_4$] (d) at (0.7,-0.7);
    \vertex (a1) at (-0.31,0.31) ;
    \vertex (b1) at ( -0.31,-0.31);
    \vertex (c1) at (0.31, 0.31);
    \vertex (d1) at (0.31,-0.31);
    \diagram* {
      (a) -- [dashed] (m) -- [dashed](b),
      (c) -- [dashed] (m) -- [dashed] (d) };
  \end{feynman}
\end{tikzpicture}}}
+ 
\vcenter{\hbox{\begin{tikzpicture}
\begin{feynman}
  every vertex/.style={red, dot};
every particle/.style={blue};
    \vertex[empty dot] (m) at (0,0) {\contour{black}{}{$G^{(6)}_{c}$}};
    \vertex[label=left:$x_1$] (a) at (-0.9,0.5) ;
    \vertex[label=left:$x_2$] (b) at ( -0.7,-0.7);
    \vertex[label=right:$x_3$] (c) at (0.9, 0.5);
    \vertex[label=right:$x_4$] (d) at (0.7,-0.7);
    \vertex (a1) at (-0.35,0.11) ;
    \vertex (b1) at ( -0.25,-0.51);
    \vertex (c1) at (0.35, 0.11);
    \vertex (d1) at (0.25,-0.51);
    \vertex (d2) at (0.15,0.47);
    \vertex (d3) at (-0.15,0.47);
    \vertex (d4) at (0,1.1);
    \diagram* {
      (a) -- [dashed] (m) -- [dashed](b),
      (c) -- [dashed] (m) -- [dashed] (d),
      (d3) -- [dashed,half left] (d4) -- [dashed, half left] (d2)} ;
  \end{feynman}
\end{tikzpicture}}} + 
\vcenter{\hbox{\begin{tikzpicture}
\begin{feynman}
  every vertex/.style={red, dot};
every particle/.style={blue};
    \vertex[empty dot] (m) at (0,0) {\contour{black}{}{$G^{(8)}_{c}$}};
    \vertex[label=left:$x_1$] (a) at (-1.0,0.5) ;
    \vertex[label=left:$x_2$] (b) at ( -1.0,-0.5);
    \vertex[label=right:$x_3$] (c) at (0.9, 0.5);
    \vertex[label=right:$x_4$] (d) at (0.9,-0.5);
    \vertex (a1) at (-0.35,0.11) ;
    \vertex (b1) at ( -0.35,-0.11);
    \vertex (c1) at (0.35, 0.11);
    \vertex (d1) at (0.35,-0.11);
    \vertex (d2) at (0.15,0.47);
    \vertex (d3) at (-0.15,0.47);
    \vertex (d4) at (0,1.1);
    \vertex (d6) at (0.15,-0.47);
    \vertex (d5) at (-0.15,-0.47);
    \vertex (d7) at (0,-1.1);
    \diagram* {
      (a) -- [dashed] (m) -- [dashed](b),
      (c) -- [dashed] (m) -- [dashed] (d),
      (d6) -- [dashed,half left] (d7) -- [dashed, half left] (d5),
      (d3) -- [dashed,half left] (d4) -- [dashed, half left] (d2)} ;
  \end{feynman}
\end{tikzpicture}}}
 + \cdots \Bigg) + O(\alpha^2), \label{eqn:feynmanedgeworth3}
\end{align}
where summing over internal points $y_i$ cancels out $\frac{1}{2n!}$ prefactor from each $G^{(2n)}_{c}$. The terms in the parenthesis constitute an infinite sum. 

This structure makes it impossible to systematically approximate $g_r(x_1, \cdots, x_r)$ with a finite number of terms via a perturbative expansion in $\alpha$, unless some other structure correlates with it. Note that this is a feature of neural network field theories where non-Gaussianities are generated \textit{only} by independence breaking. Approximation via a finite number of terms would be possible in cases where connected correlation functions scale with both $\alpha$ \textit{and} $1/N$. In the limit of $N \rightarrow \infty$, the leading-order in $\alpha$ action for a NN-FT is 
\begin{equation}
S = S_G + \sum_{r=4}^{\infty } \int d^dx_1\dots d^dx_r\, g_r(x_1,\dots,x_r) \, \phi(x_1)\dots\phi(x_r) + O(\alpha^2), \label{enq:noniidEdgeworthFeynman}
\end{equation}
where $g_{r>4}$'s are computed similar to \eqref{eqn:feynmanedgeworth3}. Such an action can not be approximated by a finite truncation, unless the theory exhibits additional structure.

\subsection{Example Actions in NN-FT \label{sec:Feynmanexamples}}

Next, we exemplify the Feynman rules from Section \eqref{sec:ContinuumEdgeworth} in a few single layer NN architecture examples at finite width and i.i.d. parameters, and evaluate the leading order in $1/N$ quartic coupling and NN-FT action. The quartic coupling is 
\begin{align}
    & g_4(x_1, \cdots , x_4)  = -\frac{1}{4!}\Big[ \int d^dy_1 \cdots d^dy_4 \,G^{(4)}_{c}(y_1, \cdots, y_4)  G_{{c}}^{(2)}(y_1, x_1)^{-1} \cdots G_{{c}}^{(2)}(y_4, x_4)^{-1} + \text{perms} \Big] , \label{eqn:EdgeFeyn1}
\end{align}
at $O(1/N)$. When $G_{{c}}^{(2)}(x_1, y_1)^{-1}$ involves differential operators, we use the methods from Appendix \eqref{app:G2inverseFT} to evaluate $g_4$.

\subsubsection*{Single Layer Cos-net}
Recall the Cos-net architecture introduced earlier, $\phi(x) = W^1_{i} \cos(W^0_{ij} x_{j} + b^0_{i})$, for $W^1 \sim \mathcal{N}(0, \sigma^2_{W_1} / N)$, $W^0 \sim \mathcal{N}(0, \sigma^2_{W_0} / d)$, and $b^0 \sim \text{Unif}[-\pi, \pi]$. We will consider the case where all parameters are independent and non-Gaussianities arise due to finite $N$ corrections. To evaluate the leading order quartic coupling for this NNFT, let us first compute the inverse propagator $G^{(2)}_{c,\text{Cos}}(x_1, x_2)^{-1}$, starting from the $2$-pt function 
\begin{align}
G^{(2)}_{c,\text{Cos}}(x_1, x_2) =\frac{\sigma_{W_1}^2}{2} e^{-\frac{\sigma_{W_0}^{2}(x_1 - x_2)^2}{2\din} },
\end{align}
and inversion relation $\int d^dy \, G^{(2)}_{c,\text{Cos}}(x, y)^{-1} \, G_{{c, \text{Cos}}}^{(2)}(y, z) = \delta^d(x-z)$. Translation invariance of the $2$-pt function and delta function constraints $G^{(2)}_{c,\text{Cos}}(x, y)^{-1}$ as a translation invariant operator. Then, performing a Fourier transformation of the $2$-pt function and its inverse operator, followed by an inverse Fourier transformation, we obtain \begin{equation}
    G^{(2)}_{c,\text{Cos}}(x, y) ^{-1}
    =\frac{2\sigma^2_{W_0}}{\sigma_{W_1}^2 \din }  \, e^{-\frac{\sigma_{W_0}^{2} \, \nabla^2_x}{2\din} } \delta^d(x-y),
\end{equation}
where $\nabla^2_x := \partial^2 / \partial x^2$. Here, we use \eqref{app:useforCosGauss} to evaluate the quartic coupling as,
\begin{align}
    g_4^{\text{Cos}}(x_1, \cdots, x_4) = & -\int d^dp_1 \cdots d^dp_4  \, \tilde{G}_{{c, \text{Cos}}}^{(4)} (p_1, \cdots , p_{4}) \, \tilde{G}_{{c, \text{Cos}}}^{(2)}(-p_1)^{-1} \cdots \tilde{G}_{{c, \text{Cos}}}^{(2)}(-p_4)^{-1} \nonumber \\
    &e^{ -i p_1x_1  \cdots  - ip_4x_4 },
\end{align}
where $\tilde{G}_{{c, \text{Cos}}}^{(4)}(p_1, \cdots, p_4)$ is from \eqref{app:cnetG4cFT}, and $\tilde{G}_{{c, \text{Cos}}}^{(2)}(-p)^{-1} = \frac{2\sigma_{W_0}}{\sqrt{\din}\sigma^2_{W_1}}e^{\frac{\din p^2}{2 \sigma^2_{W_0}}}$. Using this,
\begin{align}
    g_4^{\text{Cos}}(x_1, x_2, x_3, x_4) =  - \frac{4\sqrt{6}\pi^{3/2}\sigma^4_{W_0}}{N\din^2 \sigma^4_{W_1}}\sum_{\mathcal{P}(abcd)} e^{- \frac{\sigma^2_{W_0}\nabla^2_{r_{abcd}}}{6\din}} + \frac{8\pi \sigma^4_{W_0}}{N\din^2 \sigma^4_{W_1}}\sum_{\mathcal{P}(ab,cd)}e^{- \frac{\sigma^2_{W_0}(\nabla^2_{r_{ab}} + \nabla^2_{r_{cd}})}{2\din}}.
\end{align}
We introduce the abbreviation $r_{abcd}:= x_a + x_b - x_c - x_d$, and $\mathcal{P}(abcd) = 12$ refers to the number of ways ordered list of indices $a,c,b,d \in \{1,2,3,4 \}$ can be chosen. Similarly, $r_{ab} := x_a - x_b$, and $\mathcal{P}(ab, cd) = 12$ is the number of ways ordered pairs $(a,c),(b,d) \in \{1,2,3,4 \}$ can be drawn.

With this, Cos-net field theory action at $O(1/N)$ is 
\begin{align}
    S_{\text{Cos}}[\phi] = &\,  \frac{2 \sigma^2_{W_0}}{\sigma_{W_1}^2 {\din}}  \int d^dx \, \phi(x) \, e^{-\frac{\sigma_{W_0}^{2} \, \nabla^2_x}{2\din} } \phi(x) \, - \int d^dx_1 \cdots d^dx_4 \Bigg[ \frac{4\sqrt{6}\pi^{3/2}\sigma^4_{W_0}}{N\din^2 \sigma^4_{W_1}}\sum_{\mathcal{P}(abcd)} e^{- \frac{\sigma^2_{W_0}\nabla^2_{r_{abcd}}}{6\din}} \nonumber \\
    &- \frac{8\pi \sigma^4_{W_0}}{N\din^2 \sigma^4_{W_1}}\sum_{\mathcal{P}(ab,cd)}e^{- \frac{\sigma^2_{W_0}(\nabla^2_{r_{ab}} + \nabla^2_{r_{cd}})}{2\din}} \Bigg]   \phi(x_1) \cdots \phi(x_4) + O(1/N^2).
\end{align}
The NNGP action is local, but the leading order quartic interaction is non-local.

\subsubsection*{Single Layer Gauss-net}

As our next example, consider the output of a single-layer Gauss-net 
\begin{align}
\phi(x) = \frac{W^{1}_{i} \exp(W^{0}_{ij} x_{j} + b^{0}_{i})}{\sqrt{\exp[2(\sigma_{b_0}^2 + \frac{\sigma^2_{W_0}}{\din}x^2 )]}},
\end{align}
for parameters drawn i.i.d. from $W^{0} \sim \mathcal{N}(0,\frac{ \sigma^2_{W_0}}{\din})$, $W^{1} \sim \mathcal{N}(0, \frac{\sigma^2_{W_1}}{N})$, and $b^{0} \sim \mathcal{N}(0, \sigma^2_{b_0})$. The propagator is identical to Cos-net field theory, and so is $G^{(2)}_{c,\text{Gauss}}(x_1, x_2)^{-1}$. We evaluate Gauss-net quartic coupling $g_4$, using \eqref{app:useforCosGauss}, and \eqref{app:gnetG4cFT} for $\tilde{G}_{{c, \text{Gauss}}}^{(4)}$, as
\begin{align}
    &g_4^{\text{Gauss}}(x_1, \cdots, x_4) = - \frac{4\sqrt{2}\,\pi^{3/2}\sigma^4_{W_0}}{\sqrt{3} N^2\din^4 \sigma^4_{W_1}}\sum_{\mathcal{P}(abcd)}\bigg[ \din^2 N + 2 \sigma^4_{W_0} - \frac{\sigma^5_{W_0}(\din - \sigma^2_{W_0}\nabla^2_{r_{abcd}})}{\din^{3/2}} \bigg] e^{- \frac{\sigma^2_{W_0}\nabla^2_{r_{abcd}}}{6\din}} \nonumber \\
    &\qquad + \frac{8\pi \sigma^4_{W_0}}{N^2\din^4 \sigma^4_{W_1}}\sum_{\mathcal{P}(ab,cd)}\bigg[\din^2 N + 6\sigma^4_{W_0} - 4\din^3 \sigma^5_{W_0} + \frac{\sigma^6_{W_0}}{\din} + \Big( \frac{2\sigma^7_{W_0}}{\din^{3/2}} - \frac{\sigma^8_{W_0}}{\din^2}\Big) (\nabla^2_{r_{ab}} + \nabla^2_{r_{cd}}) \nonumber \\
    &\qquad + \frac{\sigma^{10}_{W_0}}{\din^3}\nabla^2_{r_{ab}}\nabla^2_{r_{cd}}\bigg]e^{- \frac{\sigma^2_{W_0}(\nabla^2_{r_{ab}} + \nabla^2_{r_{cd}})}{2\din}},
\end{align}
where $\mathcal{P}(ab,cd)$ and $\mathcal{P}(abcd)$ are defined as before.

Thus, Gauss-net field theory action at $O(1/N)$,  
\begin{align}
    &S_{\text{Gauss}}[\phi] = \, \frac{2 \sigma^2_{W_0}}{\sigma_{W_1}^2 {\din}}  \int d^dx \, \phi(x) \, e^{-\frac{\sigma_{W_0}^{2} \, \nabla^2_x}{2\din} } \phi(x) \, + \int d^dx_1 \cdots d^dx_4 \, g_4^{\text{Gauss}}\,\phi(x_1) \cdots \phi(x_4),
\end{align}
differs from Cos-net field theory at the level of quartic interaction.

\section{Engineering Actions: Generalities, Locality, and $\phi^4$ Theory \label{sec:designing actions}}

In Section \ref{sec:actions} we used the Edgeworth expansion and a ``duality" between fields and sources to compute couplings (including non-local ones) in the action as connected Feynman diagrams whose vertices are given by the usual connected correlators $G^{(n)}_{c}(x_1,\dots,x_n)$. This general field theory result is applicable in NN-FT of fixed architectures, but it doesn't answer the question of how to engineer an architecture that realizes a given action.

In this section we study how to design actions of a given type by deforming a Gaussian theory by an arbitrary operator. The result is simple and exploits the duality between the parameter-space and function-space descriptions of a field theory. The main results are: 
\begin{itemize}
    \item \textbf{Action Deformations.} We develop a mechanism for expressing an arbitrary deformation of a Gaussian action as a deformation of the parameter density of a NN-FT.
    \item \textbf{Local Lagrangians.} We utilize the mechanism to engineer local interactions.
    \item \textbf{$\phi^4$ Theory as a NN-FT.} Using a previous result that achieves free scalar field theory as a NN-FT, we engineer local $\phi^4$ theory as an NN-FT.
    \item \textbf{Cluster Decomposition.} We develop an approach to cluster decomposition, another notion of locality that is weaker than local interactions.
\end{itemize}
We also discuss why it might have been expected that $\phi^4$ theory (and other well-studied field theories) arises naturally at infinite-$N$.

\bigskip
To begin our analysis, consider the partition function of a Gaussian theory
\begin{equation}
Z_G[J] = \bE_G[e^{\int d^d x\, J(x) \phi(x)}],
\end{equation}
where we have labelled both the partition function and the expectation with a $G$ subscript to emphasize Gaussianity.

Now we wish to define a deformed theory that differs from the original only by an operator insertion, treating it in both function space and parameter space. The deformed partition function is given by
\begin{equation}
Z[J] = \bE_G[e^{-\lambda \int d^d x_1\dots d^d x_r \, \mathcal{O}_\phi(x_1,\dots,x_r)} e^{\int d^d x\, J(x) \phi(x)}],
\end{equation}
where $\mathcal{O_\phi}$ is a non-local operator (though it may be chosen to be local) that has a subscript $\phi$, denoting that it may depend on $\phi$ and its derivatives.
In the function space, the partition function of the Gaussian theory is
\begin{equation}
Z_G[J] = \int D\phi \, e^{-S_G[\phi] + \int d^dx J(x) \phi(x)},
\label{eqn:fnspaceZGJ}
\end{equation}
and the operator insertion corresponds to a deformation of the partition function to
\begin{equation}
Z[J] = \int D\phi \, e^{-S[\phi] + \int d^dx J(x) \phi(x)}
\label{eqn:fnspaceZJ}
\end{equation}
where the  action has been deformed 
\begin{equation}
S_G[\phi] \to S[\phi] = S_G[\phi] + \lambda \int d^d x_1\dots d^d x_r \, \mathcal{O}_\phi(x_1,\dots,x_r).
\end{equation}
We may treat this theory in perturbation theory in the usual way: correlators in the non-Gaussian theory are expanded perturbatively in $\lambda$ and evaluated using the Gaussian expectation $\bE_G$, which utilizes the Gaussian action when expressed in function-space.

How is this deformation expressed in parameter space, i.e., how do we think of this deformation from a neural network perspective? In parameter space, the Gaussian partition function is 
\begin{equation}
Z_G[J] = \int d\theta P_G(\theta) \, e^{\int d^dx J(x) \phi_\theta(x)},
\label{eqn:paramspaceZGJ}
\end{equation}
We remind the reader that in such a case Gaussianity is not obvious, but requires a judicious choice of parameter density $P(\theta)$ and architecture $\phi_\theta(x)$ such that we have a neural network Gaussian process via the central limit theorem. In parameter space, the deformation yields
\begin{equation}
Z[J] = \int d\theta P_G(\theta) \, \, e^{-\lambda \int d^d x_1\dots d^d x_r \, \mathcal{O}_{\phi_\theta}(x_1,\dots,x_r)}e^{\int d^dx J(x) \phi_\theta(x)},
\label{eqn:paramspaceZJ}
\end{equation}
where we assume that 
where the operator $\cO_{\phi_\theta}$ doesn't involve an explicit $\phi(x)$, but instead its parameter space representation; we will exemplify this momentarily. Again, correlators may be computed in perturbation theory in $\lambda$ by expanding and evaluating in the Gaussian expectation, this time in the parameter space formulation.

We emphasize that if the function space and parameter space descriptions \eqref{eqn:fnspaceZGJ} and \eqref{eqn:paramspaceZGJ} represent the same partition function, then the deformed theories \eqref{eqn:fnspaceZJ} and \eqref{eqn:paramspaceZJ} are the same theory. That is, we see how an arbitrary deformation of the action induces an associated deformation of the parameter space description. We will use this in Section \ref{sec:phi4} to engineer $\phi^4$ theory as a neural network field theory, and in \ref{sec:def_of_NNGP} we will more explicitly deform a neural network Gaussian process.

\bigskip
We end our general discussion with some theoretical considerations in neural network field theory, interpreting a non-Gaussian deformation $\cO_{\phi_\theta}$ in terms of the framework of Section \ref{sec:NG}, and also taking into account the universal approximation theorem.

A non-Gaussian deformation $\cO_{\phi_\theta}$ must violate an assumption of the CLT. The architecture itself is still the same $\phi_\theta(x)$ as in the Gaussian theory. Instead, in \eqref{eqn:paramspaceZJ} we may interpret the operator 
insertion as 
\begin{equation}
P(\theta) := P_G(\theta) \, \, e^{-\lambda \int d^d x_1\dots d^d x_r \, \mathcal{O}_{\phi_\theta}(x_1,\dots,x_r)},
\end{equation}
i.e., same architecture, but with a deformed parameter distribution. This makes it clear that our non-Gaussian theory is still at infinite-$N$ and therefore cannot receive non-Gaussianities in $1/N$-corrections. Instead, it receives non-Gaussianities because the deformed parameter distribution has independence breaking via the non-trivial relationship amongst the parameters in the deformation. There may also exist schemes for controlling non-Gaussian deformations in $1/N$, instead of via independence breaking, but it is beyond our scope.

Was it inevitable that systematic control over non-Gaussianities arises most naturally via independence breaking rather than $1/N$-corrections? The general answer is not clear, but we may use the control over non-Gaussianities to yield common theories, such as $\phi^4$ theory in the next section. In that context we may ask a related question: was it inevitable that we obtain common interacting theories via independence breaking rather than $1/N$ corrections? This question has a better answer. Finite action configurations of a common theory, say $\phi^4$ theory
\begin{equation}
S[\phi] = \int d^dx \,\left[ \phi(x)(\nabla^2 + m^2) \phi(x) + \frac{\lambda}{4!}\, \phi(x)^4\right],
\end{equation}
are not \emph{arbitrary} functions, since there may be some functions $\phi(x)$ that have infinite action. However, finite action configurations are still fairly general functions, and since they have finite action they occur with non-zero probability in the ensemble.

On the other hand, there are universal approximation theorems for neural networks, where the error in the approximation to a target function may decrease with increasing $N$. In such a case this theorem that is usually cited as a feature in ML may actually be a bug: at finite-$N$ there exist functions that can't be explicitly realized by a fixed architecture, but only approximated. We therefore find it reasonable to expect that there is at least one finite-action configuration $\phi(x)$ in $\phi^4$ theory that cannot be realized by a finite-$N$ neural network of fixed architecture; in such a case, a NN-FT realization of $\phi^4$ theory must be at infinite-$N$.
This comment only scratches the surface, but we find the interplay between universal approximation theorems and realizable field theories at finite-$N$ to be worthy of further study.

\subsection{Non-Gaussian Deformation of a Neural Network Gaussian Process \label{sec:def_of_NNGP}}
To make the general picture more concrete, we would like to consider non-Gaussian deformations of any neural network Gaussian process. The main result is that we may deform any NNGP by any operator we like, which breaks independence by deforming the parameter density, explaining the origin of non-Gaussianities by violating the independence.

As before, we consider a field built out of neurons,
\begin{equation}
\phi_\theta(x) = \frac{1}{\sqrt{N}}\sum_{i=1}^N a_i h_i(x)
\end{equation}
where the full set of parameters $\theta$ is realized by the set of parameters $a_i$ and the set of parameters $\theta_h$ of the post-activations or neurons $h$. This equation forms the field out of a linear output layer with weights $a_i$ acting on the post-activations, which could themselves be considered as the $N$-dimensional output of literally any neural network. If the reader wishes, one may take $\phi$ to be a single-layer network by further choosing 
\begin{equation}
h_i(x) = \sigma(b_{ij} x_j + c_i)
\end{equation}
with $\sigma:\bR \to \bR$ a non-linear activation function such as ReLU or tanh; with this additional choice we now have $\theta_h$ comprised of $b$-parameters and $c$-parameters. Taking the parameter densities $P_G(a)$ and $P_G(\theta_h)$ to be independent and $N\to\infty$, $\phi(x) = \phi_\theta(x)$ is drawn from a Gaussian process; we have again used a subscript $G$ to emphasize that these are the parameter densities of the Gaussian theory.

Deforming the Gaussian theory by an operator insertion, which in general is non-Gaussian, we have 
\begin{equation}
Z[J] = \int da \,d\theta_h\,\, P_G(a) P_G(\theta_h) \, \, e^{-\lambda \int d^d x_1\dots d^d x_r \, \mathcal{O}_{\phi_{a,\theta_h}}(x_1,\dots,x_r)}e^{\int d^dx J(x) \phi_\theta(x)}.
\label{eqn:nngpparamspaceZJinsertion}
\end{equation}
We may interpret the operator insertion as deforming the independent Gaussian parameter density $P_G(a) P_G(\theta_h)$ to a non-trivial joint density 
\begin{equation}
P(a,\theta_h) = P_G(a) P_G(\theta_h)\,e^{-\lambda \int d^d x_1\dots d^d x_r \, \mathcal{O}_{\phi_{a,\theta_h}}(x_1,\dots,x_r)}.
\end{equation}
The partition function is then
\begin{equation}
Z[J] = \int da \,d\theta_h\,\, P(a,\theta_h)\, e^{\int d^dx J(x) \phi_\theta(x)},
\label{eqn:nngpparamspaceZJinsertion}
\end{equation}
an infinite-$N$ non-Gaussian NN-FT where the operator insertion deforms the parameter density. At initialization, if one draws the parameters $\theta_h$ first, one may think of this as affecting the density from which the $a$-parameters are drawn; the draws of $a$-parameters are no longer independent.

For the sake of concreteness, consider the case of the single-layer network and take a general non-local quartic deformation. Then the operator insertion is 
\begin{equation}
e^{- \int d^d x_1\dots d^d x_4 \, g_4(x_1,\dots x_4) \, \phi_{a,b,c}(x_1)\dots\phi_{a,b,c}(x_4)},
\end{equation}
where Einstein summation is implied and we have absorbed the overall $\lambda$ into the definition of the non-local coupling $g_4(x_1,\dots,x_4)$. 
Inserting the equation for the neural network
\begin{equation}
\phi_{a,b,c}(x) = \frac{1}{\sqrt{N}} a_i \sigma(b_{ij} x_j + c_i),
\end{equation}
into the deformation, we obtain 
\begin{equation}
e^{- \int d^d x_1\dots d^d x_4 \, g_4(x_1,\dots x_4) \, a_{i_1}\dots a_{i_4} \sigma(b_{i_1j_1}x_{j_1} + c_{i_1})\dots \sigma(b_{i_4j_4}x_{j_4} + c_{i_4})/N^2},
\end{equation}
which defines a deformed parameter density 
\begin{equation}
P(a,b,c) = P_G(a) P_G(b) P_G(c) \, e^{- \int d^d x_1\dots d^d x_4 \, g_4(x_1,\dots x_4) \, a_{i_1}\dots a_{i_4} \sigma(b_{i_1j_1}x_{j_1} + c_{i_1})\dots \sigma(b_{i_4j_4}x_{j_4} + c_{i_4})/N^2}.
\end{equation}
Then
\begin{equation}
Z[J] = \int da \, db \, dc \,\,\,P(a,b,c)\,  e^{\int d^dx J(x)\, a_i \sigma(b_{ij} x_j + c_i) / \sqrt{N}}
\end{equation}
is the partition function of a infinite-$N$ NN-FT, as we impose $\lim N \to \infty$, with quartic non-Gaussianity induced by the breaking of independence in the joint parameter density $P(a,b,c)$.

\subsection{$\phi^4$ Theory as a Neural Network Field Theory \label{sec:phi4}}

To end this section and demonstrate the power of this technique, we would like to engineer the first interacting theory that any student learns: local $\phi^4$ theory. The action is 
\begin{equation}
S[\phi] = \int d^dx \,\left[ \phi(x)(\nabla^2 + m^2) \phi(x) + \frac{\lambda}{4!}\, \phi(x)^4\right].
\end{equation}
Following our prescription, we 
\begin{itemize}
\item \textbf{Engineer the NNGP}. Using the result of \cite{halverson2021building}, we take 
\begin{equation}
\phi_{a,b,c}(x) = \sum_i \frac{a_i \,\cos(b_{ij} x_j + c_i)}{\sqrt{\textbf{b}_i^2 + m^2}},
\end{equation}
where the sum runs from $1$ to $N=\infty$, $\textbf{b}_i$ is the vector that is the $i^\text{th}$ row of the matrix $b_{ij}$, and the parameter densities of the Gaussian theory are
\begin{align}
P_G(a) & = \prod_i e^{-\frac{N}{2\sigma_a^2} a_i a_i} \\ P_G(b) &= \prod_i P_G(\textbf{b}_i) \,\,\, \text{with} \,\,\, P_G(\textbf{b}_i)= \text{Unif}(B^d_\Lambda)\\ 
P_G(c) &= \prod_i P_G(c_i)  \,\,\, \text{with} \,\,\, P_G(c_i)=\text{Unif}([-\pi,\pi]),
\end{align}
where $B^d_\Lambda$ is a $d$-sphere of radius $\Lambda$. The density $P_G(\textbf{b}_i)$ is not independent in the vector index $j$, but all that is needed for Gaussianity is independence in the $i$ index, which is clear due to the product nature of $P_G(b)$. The power spectrum (Fourier-transform of the two-point function) is 
\begin{equation}
G^{(2)}(p) = \frac{\sigma_a^2(2\pi)^d}{2\, \text{vol}(B^d_\Lambda)}\frac{1}{p^2+m^2},
\end{equation}
which becomes the standard free scalar result $1/(p^2+m^2)$ by a trivial rescaling
\begin{equation}
\phi_{a,b,c}(x) = \sqrt{\frac{2\, \text{vol}(B^d_\Lambda)}{\sigma_a^2(2\pi)^d}}\,\,\,\sum_{i,j} \frac{a_i \,\cos(b_{ij} x_j + c_i)}{\sqrt{\textbf{b}_i^2 + m^2}}.
\label{eqn:freescalarNNGP}
\end{equation}
This neural network Gaussian process is equivalent to the free scalar theory of mass $m$ in $d$ Euclidean dimensions, with
\begin{equation}
G^{(2)}(p) = \frac{1}{p^2+m^2},
\end{equation}
where $\Lambda$ plays the role of a hard UV cutoff on the momentum.
\item \textbf{Introduce the Operator Insertion}. Given the NNGP above, or any other NNGP realizing the free scalar field theory, we wish to insert the operator 
\begin{equation}
e^{-\frac{\lambda}{4!} \int d^d x\, \phi_{a,b,c}(x)^4},
\end{equation}
associated to a local $\phi^4$ interaction.
\item \textbf{Absorb the Operator into a Parameter Density Deformation}.
The non-Gaussian operator insertion deforms the parameter density to
\begin{equation}
P(a,b,c) = P_G(a) P_G(b) P_G(c)\, \,e^{-\frac{\lambda}{4!} \int d^d x\, \phi_{a,b,c}(x)^4},\label{eqn:phi4jointparameterdensity}
\end{equation}
where for $\phi_{a,b,c}(x)$ it is to be understood that the RHS of \eqref{eqn:freescalarNNGP}
is inserted, yielding an expression that is only a function of $a$'s, $b$'s, and $c$'s.
\item \textbf{Write the Partition Function}.
We then have a partition function for the deformed theory, given by 
\begin{equation}
Z[J] = \int da \, db \, dc \,\,\,P(a,b,c)\,\,  e^{\int d^dx J(x)\, \phi_{a,b,c}(x)}, \label{eqn:phi4ZJ}
\end{equation}
where again it is to be understood that we insert the RHS of \eqref{eqn:freescalarNNGP} for $\phi_{a,b,c}$ and  \eqref{eqn:phi4jointparameterdensity} for $P(a,b,c)$; there are no explicit fields in the expression, it depends only on the architecture (which includes parameters a,b,c) and the joint parameter density.
\end{itemize}
Thus, the architecture \eqref{eqn:freescalarNNGP} and parameter density \eqref{eqn:phi4jointparameterdensity} realize local $\phi^4$ theory via the partition function
\eqref{eqn:phi4ZJ}. 
We discuss the connections between Gaussian Processes, locality, and translation invariance in Appendix \eqref{app:localityTinv}. 

Let us briefly address RG flows. The definition of a fixed non-Gaussian theory here involves the choice of a fixed value of $\lambda$, in addition to the choice of a fixed value of $\Lambda$ that was implicit in the fixing of the GP. From that starting point, decreasing $\Lambda$ while keeping the correlators fixed induces an RG flow for $\lambda$ governed by the usual Callan-Symanzik equation. In the language of the neural network architecture, this is interpreted as a flow in the parameter density that is necessary to fix the correlators as $\Lambda$ is decreased.

\subsection{Cluster Decomposition and Space Independence Breaking \label{subsec:cluster}}

We now turn to a weaker notion of locality: cluster decomposition. 
Given a field $\phi(x)$ (or neural network in our context) we say that it satisfies cluster decomposition if all connected correlation functions $G^{(r)}_{c}(x_1, \dots, x_r)$ asymptote to zero in the limit where the separation between any two space points $x_i, x_j, i \neq j$ is taken to $\infty$,
\begin{equation}
    \lim_{|x_i-x_j| \to \infty} G^{(r)}_{c}(x_1, \dots, x_r) = 0.
\end{equation}
If the probability density function of $\phi$ has the form
\begin{equation}
    P(\phi) = \frac{1}{Z} \exp{\Bigl( -\int dx \,\mathcal{L}\Bigl(x, \phi(x), \frac{\partial \phi}{\partial  x}, \dots, \frac{\partial^n \phi}{\partial x^n} \Bigr) \Bigr)}
\end{equation}
where $Z$ is a normalization constant and $n$ is finite, we say that $\phi(x)$ has a local Lagrangian density. This is a stronger notion of locality compared to cluster decomposition, as any theory with a local Lagrangian density satisfies cluster decomposition, but the converse is not true \cite{weinberg_1995}.

Checking whether a theory satisfies cluster decomposition requires knowledge of the asymptotic behavior of correlation functions, but not the probability density function. As calculating the probability density function of an NN-FT is more challenging than computing the correlation functions, checking cluster decomposition is easier than determining whether there exists a local Lagrangian density that describes the system.

The main result we describe in this section is a framework that enables engineering neural network architectures that satisfy cluster decomposition.

\subsection*{Space Independent Field Theory}
We will perform our analysis by studying, and then moving away from, a case with a very strong assumption: field theories that are defined by fields that have independent statistics at different space (or space) points $x_I$. We call these field theories \emph{space independent} (SI) field theories. While one can still view such fields as random functions defined on a continuously differentiable space, in general the field configurations are discontinuous; avoiding this would require statistical correlations between nearest neighbors, violating the assumption. This ``$d$-dimensional" field theory is really a collection of uncountably many independent $0$-d theories.
 This means that the partition function factorizes
\begin{equation} \label{eqn:ZSI}
Z_{\phi_\text{SI}}[J] = \bE[e^{\int d^d x_I J(x_I) \phi(x_I)}] = \prod_I \bE_{\phi_\text{SI}}[e^{J(x_I)\phi(x_I)}],
\end{equation}
where the product runs over all space points $x_I$. This form is agnostic about the origin of the statistics and may be specified in either the function space or parameter space description. 
 In parameter space, independent statistics at different space points $x_I$ means that the SI theory has partition function
\begin{equation}
    Z_{\phi_\text{SI}}[J] = \prod_i \int d\theta_I \,P_I(\theta_I)\, e^{J(x_I) \, \phi_{\theta_I}(x_I)},
\end{equation}
i.e. each space point $x_I$ has its own ensemble of neural networks $\phi_{\theta_I}(x_I)$ with its own set of parameters $\theta_I$ that is independent of $\theta_J$ for $I\neq J$.
In function space, independence means that 
\begin{equation}
Z_{\phi_\text{SI}}[J] = \prod_I \int D\phi_I \,e^{-S[\phi(X_I)]+ {J(x_I) \phi(x_I)}},
\end{equation}
i.e., the action is such that the path integral factorizes. An immediate consequence of this factorization is that the action cannot contain derivatives of $\phi(x_I)$, as these would depend on the value of $\phi$ not only at point $x_I$, but a local neighborhood around it. Then, the action is of the form,
\begin{equation}
S(\phi(x_I)) = V[\phi(x_I)],
\end{equation}
which, turning the product into a sum in the exponent, gives the more canonical form
\begin{equation} \label{eqn:ULpartition}
    Z{\phi_\text{SI}}[J] = \int \left(\prod_I D\phi_I\right) \,e^{-\int d^d x_I \,(V[\phi(X_I)]- {J(x_I) \phi(x_I))}}.
\end{equation}
This is a field theory with a potential, but no derivatives. The field values at different points of space are independent random variables. If they are identically distributed $V[\phi(x_I)]$ is  fixed $\forall I$ and the different factors in $Z_{\text{SI}}[J]$ enjoy an $S_L$ permutation symmetry, where the number of space points $L$ is infinite in the continuum limit.

Before introducing correlations between the field values at different space points, let us first study the statistics of the SI theory. Denote the cumulants of $\phi$ at a given point $x_I$ as $\kappa_r^\phi(x_I)$. For simplicity, we will assume that the field values at different space points are identically distributed, i.e. $\kappa_r^\phi(x_I)=\kappa_r^\phi$ is fixed for all $I$, which will also be important for translation invariance. We also assume that they are mean free, $\kappa_1^\phi=0$. Next, we consider the cumulant generating functional, which takes the form
\begin{align}
    W_{\phi_{SI}} [J] &= \log \Bigl( Z_{\phi_{SI}} [J] \Bigr)
    = \log \Bigl( \prod_I \bE_{\phi_\text{SI}}[e^{J(x_I)\phi(x_I)}] \Bigr), \nonumber \\ 
    &= \int dx \log \Bigl( \bE_{\phi_\text{SI}}[e^{J(x)\phi(x)}] \Bigr), \nonumber \\
    &= \int dx \, W[J; x],
    \label{eqn:phiCGF}
\end{align}
where $W[J; x]$ is the CGF of $\phi$ at space point $x$. Just as the partition function $Z[J]$ factorizes into a product of partition functions associated to individual space points, the CGF $W[J] = \log Z[J]$ becomes a sum (or integral, in this case). The connected correlators are easily computed by taking derivatives \footnote{We remind the reader that space derivatives are ill-defined, as $\phi(x)$ is discontinuous everywhere. However, derivatives with respect to $J(x_I)$ are still well defined. }
, where $\partial J(x_I)/\partial J(x_J) = \delta (x_I-x_J)$,
\begin{align}
    G^{(n)}_{c }  (x_1, \ldots, x_n) &= 
    \Bigl( \prod_{I=1}^n \frac{\partial}{\partial J(x_I)} \Bigr) W_{\phi_{SI}} [J], \nonumber \\
    &= \int dx \Bigl( \prod_{I=1}^n \frac{\partial}{\partial J(x_I)} \Bigr)  W [J; x].
\end{align}
and the connected correlation functions of SI networks $\phi_{\text{SI}}$ simplifies to
\begin{align}
    G^{(n)}_{c }  (x_1, \ldots, x_n) &= 
    \int dx \Bigl(\frac{\partial}{\partial J(x)}\Bigr)^n W [J; x] \prod_{I=1}^n \delta(x-x_I), \nonumber \\
    &= \int dx \, \kappa_n^{\phi} \prod_{I=1}^n \delta(x-x_I),
    \label{eqn:phiConnCorr}
\end{align}
with $n$ delta functions. The $n$-point connected correlator is nonzero only when $x_1=x_2=\dots=x_n$, and its magnitude is determined by $\kappa_n^{\phi}$. 

The correlation functions can be written in terms of the connected correlators. For example, the two point function of $\phi$ is
\begin{align}
    G^{(2)} (x_1, x_2) &= \mathbb{E}_{\phi} [\phi(x_1) \phi(x_2)], \nonumber \\
    &= \kappa_2^\phi \, \delta(x_1-x_2) + (\kappa_1^\phi)^2 \nonumber \\
    &= \kappa_2^\phi \, \delta(x_1-x_2).
\end{align}
As $\phi(x_1)$ and $\phi(x_2)$ are independent and mean free, $G^{(2)}_{\phi_{\text{SI}}} (x_1, x_2)$ is nonzero only when $x_1=x_2$. Similarly, the four point function is
\begin{align}
    G^{(4)}(x_1, x_2, x_3, x_4) &= \kappa_4^\phi \, \delta(x_1-x_2) \delta(x_1-x_3) \delta(x_1-x_4) 
    + (\kappa_2^\phi)^2 \Big(\delta(x_1-x_2) \delta(x_3-x_4) \nonumber \\
    &+ \delta(x_1-x_3) \delta(x_2-x_4) + \delta(x_1-x_4) \delta(x_2-x_3)\Big).
\end{align}
The statistics of the theory is completely determined by the space independence assumption and the cumulants $\kappa_r^\phi$. 
The general $n$-point function can be expressed as
\begin{align}
    G^{(n)} (x_1, \ldots, x_n) = \sum_{\alpha \in S_n} \prod_{r \in \alpha} G^{(r)}_{\phi_{\text{SI}}, c }  (x_1, \ldots, x_n),
    \label{eqn:phiGeneralCorr}
\end{align}
where $S_n$ denotes partitions of the set $\{1,\ldots,n\}$.

\subsection*{Space-Time Independence Breaking \label{sec:locality_ind_breaking}}
Clearly we don't want to stop with space independent theories.
We will now introduce correlations between different space points to `stitch together' the $L$ 0-dimensional theories (associated to the space independent fields) into a $d$-dimensional field theory. This requires modifying the theory in some way so that there are non-trivial correlations between field values at different points.

One way to do so is to define new field variables $\Phi(x_I)$ as a function of the SI fields $\phi(x_I)$,
\be
\Phi(x_I)=\Phi\Big(\phi(x_1), \ldots , \phi(x_L)\Big).
\label{eqn:Phifunction}
\ee
As the value of $\Phi$ at site $x_I$ in principle depends on the values of $\phi$ at all space points, $\Phi(x_I)$ and $\Phi(x_J)$ are correlated in general, even when $I \neq J$. The statistics of $\Phi(x_I)$ are then determined by the functional form of \eqref{eqn:Phifunction}, as well as the statistics of $\phi(x_I)$. However,
such a general formulation \eqref{eqn:Phifunction} is unwieldy, and we therefore simplify the picture.

We will describe a family of architectures where $\Phi(x_I)$ is constructed by a simpler ans\"atz, a smearing  of $\phi(a)_{a \in \{x_1, \cdots , x_L \}}$ across all space points, and write down a necessary and sufficient condition to satisfy cluster decomposition. Consider the architecture,
\be
\Phi(x_I) = \int_{-\infty}^{\infty} da \, f(x_I-a) \, \phi(a)
\ee
for some continuous and differentiable function $f(x_I-a)$.
First, note that although a generic draw of $\phi(a)$ is discontinuous due to independence across different points in space, $\Phi(x_I)$ is rendered continuous by the smearing. 
Furthermore, if the function $f$ is nonzero everywhere, $\Phi(x_I)$ will have correlations between all pairs of lattice sites.

We wish to check whether cluster decomposition is satisfied, and therefore need to compute correlation functions of $\Phi(x)$. The $\Phi$-correlators are given by
\begin{align}
G^{(n)}_{\Phi} (x_1, \ldots, x_n) &= \mathbb{E}_{\phi}\big[\Phi(x_1) \cdots \Phi(x_n) \big] \nonumber , \\ 
&= \mathbb{E}_{\phi}\Big[ \prod_{i=1}^n \int d a_i \, f(x_i-a_i) \, \phi(a_i) \Big].
\end{align}
As $f$ does not depend on $\phi$, we can carry out the expectation value over $\phi$ to obtain
\begin{align}
    G^{(n)}_{\Phi} (x_1, \ldots, x_n) &= \int \prod_{i=1}^n d a_i \, f(x_i-a_i) \,G^{(n)}_{\phi} (a_1, \ldots, a_n),
    \label{eqn:SmearedCorr}
\end{align}
where $G^{(n)}_{\phi} (a_1, \ldots, a_n)$ is the $n$-point correlation function of $\phi$. The only contribution to the connected correlator of $\Phi(x)$ comes from the connected piece of $G^{(n)}_{\phi} (a_1, \ldots, a_n)$ with $n$ delta functions\footnote{The remaining terms factorize and do not contribute to the connected correlator.},
\begin{equation}
G^{(n)}_{c} (x_1, \ldots, x_n) = \kappa_n^{\phi}  \int dx \prod_{i=1}^n da_i \, f(x_i-a_i) \,  \delta(x-a_i).
\end{equation}
Evaluating the integral, we obtain
\begin{equation} \label{eqn:PhiConnCorr}
   G^{(n)}_{c} (x_1, \ldots, x_n) =\kappa_n^{\phi} \int d x \, \prod_i^n f(x_i-x).
\end{equation}
Cluster decomposition is satisfied if and only if \eqref{eqn:PhiConnCorr} asymptotes to zero in the limit where the separation between any two of the space points $x_I, x_J$ is taken to $\infty$. Any smearing function $f(x)$ that decays faster than $1/x$ asymptotically satisfies this condition.\footnote{Note that the SI theory automatically satisfies cluster decomposition as the connected correlator, c.f. \eqref{eqn:phiConnCorr}, vanishes unless all space points coincide.}

\subsubsection*{Example: Gaussian Smearing}

We now present an example with a particular choice of the smearing function $f$ and show that the resulting theory satisfies cluster decomposition. Let
\begin{align}
    f(x) &= e^{-\frac{x^2}{\beta}}, \\
    \Phi(x) &= \int da \, e^{-\frac{(x-a)^2}{\beta}} \phi(a),
\end{align}
for some $\beta>0$. As before, we will consider a case where $\phi(x)$ at different space points are identically distributed, with cumulants $\kappa_\phi^n$.\footnote{As $\phi(x)$ are identically distributed, the cumulants do not depend on the space coordinates $x$.} Following equation \eqref{eqn:PhiConnCorr}, the cumulants of $\Phi(x)$ are then given by
\begin{align}\label{eqn:smearingexamplegcon}
    G^{(n)}_{c} (x_1, \ldots, x_n) &= \kappa_n^{\phi} \int d x \, \prod_{i=1}^n e^{-\frac{(x_i-x)^2}{\beta}}, \nonumber \\
     &= \kappa_n^{\phi} \sqrt{\frac{\pi \beta}{n}} \exp \Bigl[ M_{ij} x_i x_j /\beta \Bigr],
\end{align}
where
\be
M_{ij} = 
\begin{cases}
    \frac{2}{n}-2,& \text{if } i=j\\
    \frac{2}{n},              & \text{otherwise}
\end{cases}
\ee
This matrix is negative semidefinite, with eigenvalues $\lambda_1=\cdots=\lambda_{n-1}=-\beta/2$, $\lambda_n=0$, and the eigenvector corresponding to $\lambda_n$ is $(1,\cdots,1)$. Consequently, the cumulant $G^{(n)}_{c} (x_1, \ldots, x_n)$ vanishes when any of the $x_i$ are taken to be large, unless they coincide $x_1=\cdots=x_n$. This theory thus satisfies cluster decomposition.

The dependence of the connected correlators \eqref{eqn:smearingexamplegcon} on the space coordinates $x_i$ is completely determined by the choice of smearing function $f$, while their magnitudes depend both on $f$ as well as the cumulants $\kappa_\phi^n$. Although our main motivation here has been to engineer neural network architectures that satisfy cluster decomposition, smearing layers offer great flexibility in manipulating the connected correlators and might be useful in designing neural networks with other desired properties.

\section{Conclusions} \label{sec:Conclusions}

In this paper we continued the development of neural network field theory (NN-FT), a new approach to field theory in which a theory is specified by a neural network architecture and a parameter density. This description enables a parameter space description of the statistics, yielding a different method for computing correlation functions. For a more detailed introduction to NN-FT, see the introduction and references therein.

We focused on three foundational aspects of NN-FT: non-Gaussianity, actions, and locality. Via the central limit theorem (CLT), many architectures admit an $N\to\infty$ limit in which the associated NN-FT is Gaussian, i.e., a generalized free field theory. In the machine learning literature, these are called neural network Gaussian processes (NNGPs).
In Section \ref{sec:NG} we demonstrated that interactions arise from parametrically violating assumptions of the CLT, yielding non-Gaussianities arising from $1/N$-corrections, as well as the breaking of statistical independence and the identicalness assumption. These interactions are apparent via parameter-space calculations of connected correlation functions, but manifest themselves as non-Gaussianities in the field density $P[\phi]=\exp(-S[\phi])$. 
In Section \ref{sec:actions} we developed a technique that allows for the action to be computed from the connected correlation functions, via connected Feynman diagrams. This is an inversion of the usual approach in field theory: we compute coupling functions in terms of connected correlators, rather than the other way around. The technique was applied to NN-FT, including an analysis involving the parametric non-Gaussianities we studied.
In Section \ref{sec:designing actions} we studied how to design architectures that realize a given action. We do so by deforming an NNGP by an operator insertion that, from a function-space perspective, corresponds to a deformation of the GP action. However, since we know the architecture we may also express the deformation in parameter space, in which case the non-Gaussianity associated to a given deformation of the action has a natural interpretation as a deformation of the neural network parameter density. That is, the interactions arise from independence breaking. We apply this technique to induce local interactions, and derive an architecture that realizes $\phi^4$ theory as an infinite NN-FT.

\bigskip \noindent
\textbf{Acknowledgements.}
We thank Sergei Gukov, Mathis Gerdes, Jessica Howard, Ro Jefferson, Gowri Kurup, Joydeep Naskar, Fabian Ruehle, Jiahua Tian, Jacob Zavatone-Veth, and Kevin Zhang for discussions.
This work is supported by the National Science Foundation
under Cooperative Agreement PHY-2019786 (The NSF
AI Institute for Artificial Intelligence and Fundamental
Interactions).
This work was performed in part at Aspen Center for Physics, which is supported by National Science Foundation grant PHY-2210452. A.M. thanks ECT* and the ExtreMe Matter Institute EMMI at GSI, Darmstadt, for support in the framework of an ECT* Workshop during which part of this work has been completed. J.H. is supported by NSF CAREER grant PHY-1848089.

\begin{appendices}
\section{Continuum Hermite Polynomials} \label{app:Hermite}
Let us first recall the definition of continuum Hermite polynomials for convenience,
\begin{equation}
    H(\phi, x_1, \cdots, x_n) = (-1)^n e^{S_G} \frac{\delta}{\delta \phi(x_{1}) } \cdots \frac{\delta}{\delta \phi(x_{n})} e^{-S_G}.
\end{equation}
Defining
\begin{equation}
    S_i = \frac{\delta S_G}{\delta \phi(x_i)}, \qquad S_{i,j} = \frac{\delta^2 S_G}{\delta \phi(x_i) \delta \phi(x_j)},
\end{equation}
the first six Hermite polynomials are,
\begin{align}
    &H_1(\phi, x_1) = S_1, \nonumber \\
    &H_2(\phi, x_1, x_2) = S_1 S_2 - S_{1,2}, \nonumber \\
    &H_3(\phi, x_1, x_2, x_3) = S_1 S_2 S_3 - S_{1,2} S_3 [3], \nonumber\\
    &H_4(\phi, x_1, x_2, x_3, x_4) = S_1 S_2 S_3 S_4 - S_{1,2} S_3 S_4 [6] + S_{1,2} S_{3,4} [3], \nonumber\\
    &H_5(\phi, x_1, x_2, x_3, x_4, x_5) = S_1 S_2 S_3 S_4 S_5 - S_{1,5} S_2 S_3 S_4 [10] + S_{1,2} S_{3,4} S_5 [15], \nonumber\\
    &H_6(\phi, x_1, x_2, x_3, x_4, x_5, x_6) = S_1 S_2 S_3 S_4 S_5 S_6 - S_{1,6} S_2 S_3 S_4 S_5 [15]  \nonumber \\ 
    &\,  \qquad \qquad  \qquad \qquad + S_{1,2} S_{3,4} S_5 S_6 [45]  - S_{1,2} S_{3,4} S_{5,6} [15],
\end{align}
where the square brackets denote sums over all terms with a given index structure, for example $S_{1,2} S_3 [3] = S_{1,2} S_3 + S_{1,3} S_2 + S_{2,3} S_1$.
\section{Details of Examples} \label{appendix_eg}

\subsection*{ReLU-net Cumulants at Finite $N$, i.i.d. Parameters} 
Let us study the output distribution of a single hidden layer network at width $N$, ReLU activation function, $\din = \dout = 1$, given by  
\begin{equation}
\phi(x) = W^1_{i} R(W^0_{ij} x_{j} )~~\text{where}~ R(z) = \begin{cases} z,~\text{for}~ z\geq 0 \\ 0,~ \text{otherwise} \end{cases} .
\end{equation}
The parameters are sampled i.i.d., $W^{0} \sim \mathcal{N}(0, \frac{ \sigma^2_{W_0} }{\din})$, $W^{1} \sim \mathcal{N}(0, \frac{ \sigma^2_{W_1} }{ N})$, and bias = $0$. The $2$-pt function is $G^{(2)}_{c,\text{ReLU}}(x,y) = \sigma^2_{W_0}\sigma^2_{W_1} \big( R(x)R(y) + R(-x)R(-y) \big)/2$, and higher order cumulants are
\begin{align}
\label{eqn:g4con}
G^{(4)}_{c,\text{ReLU}}(x_1, \cdots, x_4) =& ~\frac{1}{N}\Bigg( \frac{15 \sigma^4_{W_0} \sigma^4_{W_1}}{4\din^2} \Big(\sum\limits_{j=\pm 1} R(j x_1)R(jx_2)R(jx_3)R(jx_4)\Big) \nonumber \\
&- \frac{\sigma^4_{W_0} \sigma^4_{W_1}}{4\din^2 }\Big( \sum\limits_{\mathcal{P}(abcd)} \sum\limits_{j= \pm 1} R(j x_a)R(j x_b)R(-j x_c)R(-j x_d) \Big) \Bigg),
\end{align}
\begin{align}
G^{(6)}_{c,\text{ReLU}}(x_1, \cdots, x_6) =  \frac{1}{N^2} \Bigg[ \frac{225\, \sigma^6_{W_0} \sigma^6_{W_1} }{2\din^3} \Big( \sum\limits_{j= \pm 1} R(jx_1)R(jx_2)R(jx_3)R(jx_4)R(jx_5)R(jx_6) \Big)&  \nonumber \\
-  \sum_{\mathcal{P}(abcdef)}\Bigg( \frac{9\, \sigma^6_{W_0} \sigma^6_{W_1} }{4\din^3} \Big(\sum\limits_{j_1 = \pm 1} R(j_1x_a)R(j_1x_b)R(j_1x_c)R(j_1x_d)\Big) \Big(\sum\limits_{j_2= \pm 1} R(j_2 x_e)R(j_2 x_f)\Big)& \nonumber \\
- \frac{\sigma^6_{W_0} \sigma^6_{W_1} }{4\din^3} \Big(\sum\limits_{j_1= \pm 1} R(j_1x_a)R(j_1x_b)\Big) \Big(\sum\limits_{j_2= \pm 1} R(j_2 x_c)R(j_2 x_d)\Big) \Big(\sum\limits_{j_3= \pm 1} R(j_3x_e)R(j_3x_f)\Big) \Bigg) \Bigg]&   ,
\end{align}
where $\mathcal{P}(abcd)$ denotes all combinations of non-identical $a,b,c,d$ drawn from $\{1,2,3,4\}$, and similarly for $\mathcal{P}(abcdef)$.

\subsection*{Cos-net Cumulants at Finite $N$, Non-i.i.d. Parameters} \label{appendix:cosnet_eg}
The output of a single hidden layer, finite $N$, fully connected feedforward network with cosine activation function is given by,
\begin{equation}
    \phi(x) = W^1_{i} \cos(W^0_{ij} x_{j} + b^0_{i}).
\end{equation}
For i.i.d. parameters, e.g. $W^1 \sim \mathcal{N}(0, \sigma^2_{W_1} / N)$, $W^0 \sim \mathcal{N}(0, \sigma^2_{W_0} / d)$, and $b^0 \sim \text{Unif}[-\pi, \pi]$, the $2$-pt function is given by $G^{(2)}_{c,\text{Cos}}(x, y) = \frac{\sigma_{W_1}^2}{2} e^{-\frac{1}{2\din}\sigma_{W_0}^{2} (x - y)^2}$. For simplicity, we focus on the $\din =1$ case; the statistical independence of first linear layer weights can be broken by a hyperparameter $\aIB \ll 1$, then the correlated weight distribution is
\begin{align}\mathcal{P}(W^0) = c \exp{\Bigg[-\sum_{i} \Big( \frac{(W^0_{i})^2}{2 \sigma_{W_0}^2}  + \frac{\alpha_{\text{IB}}}{N^2} \sum_{i_1, i_2 } (W^0_{i_1})^2 (W^0_{i_2})^2 \Big) \Bigg]},
\end{align}
where $c$ is a normalization constant. The cumulative non-Gaussian effects due to finite width and non-i.i.d. parameters alter all correlation functions, including the $2$-pt function at finite width. Using perturbation theory at leading order in $\aIB$, the $2^{\text{nd}}$ and $4^{\text{th}}$ cumulants are evaluated as the following,
\begin{equation}
\begin{split}
G^{(2)}_{c,\text{Cos}}(x_1, x_2) =&\,  \frac{\aIB  \sigma_{W_0}^4 \sigma_{W_1}^2 e^{-\frac{\sigma_{W_0}^2 (\Delta x_{12})^2}{2 }} }{2 N}\Big[ \left(1+\sigma_{W_0}^2 (\Delta x_{12})^2\right)  \\ &- \frac{\left(1-5  \sigma_{W_0}^2 (\Delta x_{12})^2+\sigma_{W_0}^4 (\Delta x_{12})^4\right)}{N} \Big] ,
\end{split}
\end{equation}
\begin{align}
\label{cnetG4NGP}
  &  G^{(4)}_{c,\text{Cos}}(x_1, x_2, x_3, x_4) = \frac{\sigma_{W_1}^4 }{8  N } \sum\limits_{\mathcal{P}(abcd)} \Bigg[ \bigg(-2 e^{-\frac{\sigma_{W_0}^2 \left((\Delta x_{ab})^2+(\Delta x_{cd})^2\right)}{2 }} +3 e^{-\frac{\sigma_{W_0}^2 (\Delta x_{ab} + \Delta x_{cd} )^2}{2 }} \bigg)
\nonumber    \\ &+ \frac{ \aIB   \sigma_{W_0}^4 }{  N} \bigg(-6 e^{-\frac{\sigma_{W_0}^2 \left((\Delta x_{ab})^2+(\Delta x_{cd})^2\right)}{2 }}+3 e^{-\frac{\sigma_{W_0}^2 (\Delta x_{ab} + \Delta x_{cd})^2}{2 }} +3  \sigma_{W_0}^2 (\Delta x_{ab} + \Delta x_{cd})^2 \nonumber  \\
    & \, e^{-\frac{\sigma_{W_0}^2 (\Delta x_{ab} + \Delta x_{cd})^2}{2 }} -2  \sigma_{W_0}^2 \left((\Delta x_{ab})^2+(\Delta x_{cd})^2\right) e^{-\frac{\sigma_{W_0}^2 \left((\Delta x_{ab})^2+(\Delta x_{cd})^2\right)}{2 }} -2 \sigma_{W_0}^4 (\Delta x_{ab})^2 (\Delta x_{cd})^2 \nonumber \\
    &\, e^{-\frac{\sigma_{W_0}^2 \left((\Delta x_{ab} )^2+(\Delta x_{cd})^2\right)}{2 }}  \bigg) \Bigg] ,
\end{align}
where $\Delta x_{ij} := x_i - x_j$. The Fourier transformation of this cumulant at $\aIB = 0$ is 
\begin{align} \label{app:cnetG4cFT}
    & \tilde{G}_{{c, \text{Cos}}}^{(4)} = \frac{3\pi^{3/2}\sigma^4_{W_1}\sqrt{\din}}{2\sqrt{2}N \sigma_{W_0}}\Bigg[ e^{-\frac{p^2_1 \din}{2\sigma^2_{W_0}}}\Big( \delta^{\din}(p_1 + p_2)\delta^{\din}(p_1 + p_3) \delta^{\din}(p_4 - p_1) + \delta^{\din}(p_2 - p_1)\delta^{\din}(p_1 + p_3)\nonumber \\
    &\delta^{\din}(p_1 + p_4) + \delta^{\din}(p_1 + p_2)\delta^{\din}(p_3 - p_1)\delta^{\din}(p_1 + p_4) \Big)\Bigg] - \frac{\pi \sigma^4_{W_1}\din}{2N\sigma^2_{W_0}} \Bigg[e^{-\frac{ ( p^2_1 + p^2_2){\din}}{2\sigma^2_{W_0}}} \delta^{\din}(p_1 + p_4)\delta^{\din}(p_2 + p_3) \nonumber \\
    & e^{-\frac{( p^2_1 + p^2_2)\din}{2\sigma^2_{W_0}}} \delta^{\din}(p_1 + p_3)\delta^{\din}(p_2 + p_4) + e^{-\frac{( p^2_1 + p^2_3)\din}{2\sigma^2_{W_0}}} \delta^{\din}(p_1 + p_2)\delta^{\din}(p_3 + p_4)\Bigg] + p_1 \leftrightarrow p_2, p_3, p_4 ,
\end{align}
where use the convention $e^{i(p_1x_1 + p_2x_2 + p_3x_3 + p_4x_4)}$.

Next, we present another example where non-Gaussianities arise due to both finite width and non-i.i.d. parameters. 

\subsection*{Gauss-net at Finite $N$, Non-i.i.d. Parameters} \label{appendix:gaussnet_eg}
We define the Gauss-net architecture as a single hidden layer, width $N$, feedforward network with exponential activation function, and an overall normalizing factor, such that the output is 
\begin{equation}
    \phi(x) = \frac{W^{1}_{i} \exp(W^{0}_{ij} x_{j} + b^{0}_{i})}{\sqrt{\exp[2(\sigma_{b_0}^2 + \frac{\sigma^2_{W_0}}{\din}x^2 )]}}.
\end{equation}
For i.i.d. parameter distributions, $W^{0} \sim \mathcal{N}(0,\frac{ \sigma^2_{W_0}}{\din})$, $W^{1} \sim \mathcal{N}(0, \frac{\sigma^2_{W_1}}{N})$, and $b^{0} \sim \mathcal{N}(0, \sigma^2_{b_0})$, the $2$-pt function is $G^{(2)}_{c,\text{Gauss}}(x, y) = \frac{\sigma_{W_1}^2}{2} e^{-\frac{1}{2\din}\sigma_{W_0}^{2} (x - y)^2}$, identical as Cos-net. We break the statistical independence of the first linear layer weights similar to the previous example, at $\din=1$. Then, the $2^{\text{nd}}$ and $4^{\text{th}}$ order cumulants at leading order in $\aIB$ are,
\begin{align}
G^{(2)}_{c,\text{Gauss}}(x_1, x_2) &=  -\frac{\aIB  \sigma_{W_0}^4 \sigma_{W_1}^2 }{2  N }  e^{-\frac{\sigma_{W_0}^2 (\Delta x_{12})^2}{2 }} \Big[ \left(\sigma_{W_0}^2 X_{12}^2-1 \right)
    -\frac{ \left(1+5  \sigma_{W_0}^2 X_{12}^2+\sigma_{W_0}^4 X_{12}^4\right)}{  N } \Big],
\end{align}
and,
\begin{align}
    & G^{(4)}_{c,\text{Gauss}}(x_1, x_2, x_3, x_4) = \frac{3 \sigma_{W_1}^4 \exp \left(-\frac{\sigma_{W_0}^2 \left(x_{1}^2-2 x_{1} (x_{2}+X_{34})+x_{2}^2-2 x_{2} X_{34} +(\Delta x_{34})^2\right)}{2 }\right)}{4  N } \nonumber \\
    & + \frac{ \aIB   \sigma_{W_0}^4 \sigma_{W_1}^4 }{4  N ^2} \Bigg(3 \exp \left(-\frac{\sigma_{W_0}^2 \left(x_{1}^2-2 x_{1} (x_{2}+X_{34})+x_{2}^2-2 x_{2}X_{34}+(\Delta x_{34})^2\right)}{2 }\right) \nonumber \\&  -3  \sigma_{W_0}^2 (X_{12}+X_{34})^2 \exp \left(-\frac{\sigma_{W_0}^2 \left(x_{1}^2-2 x_{1} (x_{2}+X_{34})+x_{2}^2-2 x_{2} X_{34} +(\Delta x_{34})^2\right)}{2 }\right) \nonumber \\& 
-  \sum\limits_{\mathcal{P}(abcd)}\bigg( 3 \, e^{-\frac{\sigma_{W_0}^2 \left((\Delta x_{ab})^2+(\Delta x_{cd})^2\right)}{2 }} - \sigma_{W_0}^2 \left(X_{ab}^2+ X_{cd}^2\right) e^{-\frac{\sigma_{W_0}^2 \left((\Delta x_{ab})^2+(\Delta x_{cd})^2\right)}{2 }} \nonumber \\
&+ \sigma_{W_0}^4 X_{ab}^2 \, X_{cd}^2 \, e^{-\frac{\sigma_{W_0}^2 \left((\Delta x_{ab})^2+(\Delta x_{cd})^2\right)}{2 }} \bigg)\Bigg) -\sum\limits_{\mathcal{P}(abcd)} \frac{\sigma_{W_1}^4}{4  N } e^{-\frac{\sigma_{W_0}^2 \left((\Delta x_{ab})^2+(\Delta x_{cd})^2\right)}{2 } }, \label{gnetG4NGP}
\end{align}
 where $X_{ij} := x_i + x_j$, and $\Delta x_{ij} := x_i - x_j$.

 At $\aIB =0$, the Fourier transformation of this cumulant becomes the following
 \begin{align} \label{app:gnetG4cFT}
     & \tilde{G}_{{c, \text{Gauss}}}^{(4)} = \frac{\pi^{3/2}\sigma^4_{W_1}}{2\sqrt{2}N^2 \din^{3/2} \sigma_{W_0}}\Bigg[ e^{-\frac{ p^2_1\din}{2\sigma^2_{W_0}}}( \din^2 N - \din p_1^2 \sigma^2_{W_0} + 2\sigma^4_{W_0}) \Big( \delta^{\din}(p_1 + p_2)\delta^{\din}(p_1 + p_3) \nonumber \\
     & \delta^{\din}(p_4 - p_1) + \delta^{\din}(p_2 - p_1)\delta^{\din}(p_1 + p_3)\delta^{\din}(p_1 + p_4) + \delta^{\din}(p_1 + p_2)\delta^{\din}(p_3 - p_1)\delta^{\din}(p_1 + p_4) \Big)\Bigg] \nonumber \\
    &- \frac{\pi \sigma^4_{W_1}}{2  N^2 \sigma^2_{W_0} \din} \Bigg[\big(\din^2(N + p_1^2 p_2^2)-2\din(p_1^2 + p_2^2)\sigma^2_{W_0} + 6\sigma^4_{W_0}  \big) \Big(e^{-\frac{( p^2_1 + p^2_2)\din}{2\sigma^2_{W_0}}} \delta^{\din}(p_1 + p_4)\delta^{\din}(p_2 + p_3) \nonumber \\
    & e^{-\frac{( p^2_1 + p^2_2)\din}{2\sigma^2_{W_0}}} \delta^{\din}(p_1 + p_3)\delta^{\din}(p_2 + p_4)\Big) + \big(\din^2 (N + p_1^2 p_3^2)-2\din (p_1^2 + p_3^2)\sigma^2_{W_0} + 6\sigma^4_{W_0}\big)e^{-\frac{( p^2_1 + p^2_3)\din}{2\sigma^2_{W_0}}} \nonumber \\
    &\delta^{\din}(p_1 + p_2)\delta^{\din}(p_3 + p_4)\Bigg] + p_1 \leftrightarrow p_2, p_3, p_4 ,
 \end{align}
 using the same convention as Cos-net.

\subsection*{Non-Gaussianity from Non-Identical Parameter Distributions} \label{appendix:noniid_ngp}

We discussed examples of NN architectures where non-Gaussianities arise at various widths, from the choice of identical but correlated parameter distributions. In addition to this, it is possible to violate CLT through independently drawn dissimilar NN parameter distributions; this too induces non-Gaussianities in NN output distributions. Let us present an architecture where non-Gaussianities at infinite width limit arise due to dissimilar \emph{and} independent parameter distributions. Consider the NN architecture with output
\begin{align}
    \phi(x_k) = \sum\limits_{j=-N}^{N}e^{- \frac{j^2}{\sigma^2}}\, W^{L}_{j} h^{L-1}_j (x_{k}) + b^{L},  
\end{align}
with parameters drawn from $W^{1} \sim \mathcal{N}(0, \sigma_{W_L}^2),\, b^{L} \sim \mathcal{N}(0, \sigma_{b_L}^2)$, and $h^{L-1}_{j}(x_k)$ denotes the output of $j^{\text{th}}$ neuron in $(L-1)^{\text{th}}$ hidden layer, from input $x_k$. The presence of the prefactor $e^{- \frac{j^2}{\sigma^2}}$ at the $j^{\text{th}}$ node of final linear layer leads to dissimilarities in the final layer parameter distributions. Let us study the first three leading order cumulants at $\lim N \to \infty$,
\begin{align}
 G^{(2)}_{c}(x_1, x_2) &= \lim_{N \rightarrow \infty} \sum\limits_{j=-N}^{N} e^{- \frac{ 2j^2}{\sigma^2}}\, \sigma^2_{W_L} 
 \mathbb{E}[h^{L-1}_j(x_1)h^{L-1}_j(x_2)] = \sqrt{\frac{\pi}{2}} \sigma \, \sigma^2_{W_L} 
 \mathbb{E}[h(x_1)h(x_2)]  \label{G2conneg1}, \\
    G^{(4)}_{c}(x_1, \cdots, x_4) &= \sqrt{\frac{\pi}{4}} \sigma \, \sigma^4_{W_L} \Big[ 3\,
 \mathbb{E}[h(x_1)\cdots h(x_4)] - \sum\limits_{\mathcal{P}(abcd)} \mathbb{E}[h(x_a)h(x_b)]\mathbb{E}[h(x_c)h(x_d)] \Big] \label{G4conneg1}, \end{align}
 and
 \begin{align}
    &G^{(6)}_{c}(x_1, x_2, x_3, x_4, x_5, x_6) = \sqrt{\frac{\pi}{6}} \sigma \, \sigma^6_{W_L} \Big[ 15\,
 \mathbb{E}[h(x_1)h(x_2)h(x_3)h(x_4)h(x_5)h(x_6)]  - 3 \sum\limits_{\mathcal{P}(abcdef)} \nonumber \\
 & \Big( \mathbb{E}[h(x_a)h(x_b)h(x_c)h(x_d)]\mathbb{E}[h(x_e)h(x_f)]  -2 \, \mathbb{E}[h(x_a)h(x_b)]\mathbb{E}[h(x_c)h(x_d)]\mathbb{E}[h(x_e)h(x_f)] \Big) \Big].\label{G6conneg1}
\end{align}
We used $h(x):= h^{L-1}(x)$, and identities $\mathbb{E}[(W^{L}_{j})^6] = 15 \, \sigma_{W_1}^6$, $\mathbb{E}[(W^{L}_{j})^4] = 3 \, \sigma_{W_1}^4$. All these cumulants are nonvanishing at $\lim N \to \infty$; similarly, one can show that other higher order cumulants are non-vanishing too, adding non-Gaussianities to the output distribution.

\section{CGF and Edgeworth Expansion for NNFT } \label{app:CGFedgeworth}
We express the output of a single hidden layer width $N$ neural network as a sum over $N$ continuous variables
\begin{align}
\phi(x) = \frac{1}{\sqrt{N}} \sum\limits_{i=1}^{N} h_i(x),
\end{align}
where $h_i(x)$ are the outputs of each neuron before they get summed up into the final output. 
\paragraph{Finite N and I.I.D. Parameters}
The cumulant generating functional for i.i.d. parameters $P(h;\vec{\alpha}=\vec{0}) = \prod\limits_{i=1}^{N} P_i(h_i)$ become the following 
\begin{align}
W_{\phi(x)}[J] =& \log \mathbb{E} \Big[e^{\frac{1}{\sqrt{N}} \sum\limits_{i=1}^{N} \int dx J(x)h_i(x)   } \Big] \nonumber \\
=& \log\Big[\prod^{N}_{i=1} \int\, Dh_i \, P_i(h_i) \, \exp\Big( \frac{1}{\sqrt{N}} \int dx J(x) h_i(x)  \Big)\Big] \nonumber \\
=& N \log\Big[ \sum_{r=0}^{\infty} \prod^{r}_{i=1} \int\, dx_i \frac{G^{(r)}_{h_i}(x_1, \cdots, x_r)J(x_1)\cdots J(x_r) }{r! N^{r/2}} \Big] \nonumber \\
=& \sum_{r=0}^{\infty}  \prod^{r}_{i=1} \int\, dx_i \frac{G^{(r)}_{c, h_i}(x_1, \cdots, x_r)J(x_1)\cdots J(x_r) }{r! N^{r/2 - 1}}
\end{align}
where $J(x)$ and $h_i(x)$ are the source current and output of $i^{\text{th}}$ neuron, respectively. In the second last step, we have used the following relation,
\begin{align}
 \sum_{r=0}^{\infty} \prod^{r}_{i=1} \int\, dx_i \frac{G^{(r)}_{h_i}(x_1, \cdots, x_r)J(x_1)\cdots J(x_r) }{r! N^{r/2}} = e^{ \,\, \sum\limits_{r=0}^{\infty} \frac{1}{r! N^{r/2}  } \int \big( \prod\limits^{r}_{i=1} dx_i \big) G^{(r)}_{c, h_i }(x_1, \cdots, x_r)J(x_1)\cdots J(x_r) } . 
 \end{align}
 Lastly, we use $W[J] = \sum\limits_{r=0}^{\infty} \big( \prod\limits^{r}_{i=1} \int\, dx_i \big) \frac{G^{(r)}_{c }(x_1, \cdots, x_r)J(x_1)\cdots J(x_r) }{r! } $ to obtain 
 \begin{align}
 G^{(r)}_{c }(x_1, \cdots, x_r)J(x_1)\cdots J(x_r) = \frac{G^{(r)}_{c, h_i}(x_1, \cdots, x_r)J(x_1)\cdots J(x_r) }{N^{r/2 - 1}},
 \end{align}
 with a $N$-scaling of cumulants, as expected. 
 
 \paragraph{Correlated Parameters at Finite $N$}
Let $\vec{\alpha} = \{\alpha_1, \cdots, \alpha_q \}$ be parameters breaking statistical independence between neurons. For small $\vec{\alpha}$, one can write
\begin{align}
P(h;\vec{\alpha}) = P(h;\vec{\alpha} = \vec{0}) + \sum\limits^{\infty}_{r=1} \sum\limits^{q}_{s_1,\cdots, s_r=1} \frac{\alpha_{s_1}\cdots \alpha_{s_r} }{r!}\partial_{\alpha_{s_1}} \cdots \partial_{\alpha_{s_r}} P(h;\vec{\alpha}) \Big|_{\vec{\alpha} = 0}\,\,.
\end{align}  
One can define the $r^{\text{th}}$ derivative as $\partial_{\alpha_{s_1}} \cdots \partial_{\alpha_{s_r}} P(h;\vec{\alpha})  = P(h;\vec{\alpha}) \mathcal{P}_{r, \{s_1 , \cdots ,s_r \} }$; the recursive relation satisfied by $\mathcal{P}_{r, \{s_1, \cdots, s_r \} }$ is 
\begin{align}
\mathcal{P}_{r +1, \{s_1, \cdots, s_{r +1} \}} = \frac{1}{r+1}\sum_{\gamma = 1}^{r + 1} (\mathcal{P}_{1, s_{\gamma}} + \partial_{\alpha_{s_{\gamma}}}) \mathcal{P}_{r, \{s_1, \cdots, s_{r +1} \}\backslash s_{\gamma} } .
\end{align}
With this, the NN parameter distribution can be expressed as 
\begin{align}
P(h;\vec{\alpha}) = P(h;\vec{\alpha} = \vec{0}) + \sum\limits^{\infty}_{r=1} \sum\limits^{q}_{s_1,\cdots, s_r=1} \frac{\alpha_{s_1}\cdots \alpha_{s_r} }{r!}  P(h;\vec{\alpha}) \mathcal{P}_{r, \{s_1 , \cdots ,s_r \} } \big|_{\vec{\alpha} = 0}
\end{align}

Next, let us derive the CGF for the NN functional distribution,
\begin{align}
& W_{\vec{\alpha}}[J] =  \log \Bigg[ \int Dh \, P(h;\vec{\alpha}) \, e^{ \, \frac{1}{\sqrt{N}}\sum\limits_{i=1}^{N} \int dx\, h_i(x) J(x) }  \Bigg] \nonumber \\
= & \log \Bigg[  \prod\limits^{N}_{i=1}  \mathbb{E}_{P_i(h_i)}\Bigg[ \Big(  1 + \sum_{r=1}^{\infty} \sum\limits_{s_1, \cdots, s_r = 1}^{q} \frac{\alpha_{s_1}\cdots \alpha_{s_r}}{r!} \mathcal{P}_{r, \{s_1 , \cdots , s_r \} } \big|_{\vec{\alpha} = 0} \Big) e^{ \, \frac{1}{\sqrt{N}}\int dx\, h_i (x) J(x) }  \Bigg] \Bigg] \nonumber \\
=& \log \Bigg[ e^{W_{\text{free}} [J] } + \sum_{r=1}^{\infty} \sum\limits_{s_1, \cdots, s_r = 1}^{q} \frac{\alpha_{s_1}\cdots \alpha_{s_r}}{r!} \prod\limits^{N}_{i=1}  \mathbb{E}_{P_i(h_i)}\Big[  e^{ \,  \frac{1}{\sqrt{N}}\int dx\, h_i(x) J(x) } \cdot \mathcal{P}_{r, \{s_1 , \cdots , s_r \} } \big|_{\vec{\alpha} = 0}  \Big] \Bigg] . \label{eq:CGFcont}
\end{align}
The last line is obtained using
\begin{align}
 \prod^{N}_{i=1}\mathbb{E}_{P_i(h_i) }\Big[e^{\frac{1}{\sqrt{N}} \int dx\, J(x) h_i(x) } \Big] & = \exp{ \Big( N \sum_{r=0}^{\infty} \int\, \prod^{r}_{i=1} dx_i \frac{G^{(r)}_{c, h_i}(x_1, \cdots, x_r)J(x_1)\cdots J(x_r) }{r! N^{r/2 - 1}} \Big) } \nonumber \\
 &= e^{W_{\text{free}} [J] }.
 \end{align}
At $\lim N \to \infty$, we obtain $W_{\text{free}}[J] =\int dx_1 dx_2  \frac{J(x_1) \, G^{(2)}_{c, h_i}(x_1, x_2) \, J(x_2) }{2}$. 

The partition function of a field theory is related to its CGF as
\begin{align}
Z[J(x)] = \mathbb{E}[e^{i \int J(x) \phi(x)}] =\prod_{i=1}^{N} \int Dh P_i(h_i)\,  e^{ \frac{i}{\sqrt{N}} \int dx J(x) \, h_i(x) }.
\end{align}
Under the transformation $J \rightarrow  i J$, the CGF becomes,
\begin{align}
W[J] = \sum\limits_{r=1}^{\infty} \int \prod\limits_{j=1}^{r} dx_j \, \frac{ i^r }{r!}G^{(r)}_{c}(x_{1}, \cdots, x_{r} ) J(x_{1}) \cdots J(x_{r})  =: \sum\limits^{\infty}_{r=1} \int \prod\limits_{j=1}^{r} dx_j\, \frac{i^r}{r!} G^{(r)}_{c}  \, J_{\underline{r}}, 
\end{align}
the inverse Fourier transform of which, up to renormalization, is
\begin{align} \label{eq:Edgeworthcont}
 P[\phi] \propto  & \int DJ \, e^{   W[J] - i \int dx J(x)\phi(x)   } \nonumber \\
 = & \int DJ \, e^{ \,\,   \sum\limits^{\infty}_{r=1} \int dx_1 \cdots dx_r \frac{ i^r}{r!}  G^{(r)}_{c}(x_1, \cdots , x_r) \,  J_{\underline{r}}\,  - \, i \int dx J(x)\phi(x)   } \nonumber \\
 = & \int DJ \, e^{\,\, \sum\limits^{\infty}_{r=3} \int dx_1 \cdots dx_r \frac{ i^r}{r!}  G^{(r)}_{c}(x_1, \cdots , x_r) \, J_{\underline{r}}    }  e^{- i  \int dx\, J(x) \phi(x)  } \,\, e^{\, \, \frac{ i \int dx_1\, G^{(1)}_{c}(x_1) \, J_{\underline{1}} }{1!} -  \frac{ \int dx_1 dx_2 \, G^{(2)}_{c}(x_1, x_2) \, J_{\underline{2}}    }{2!} }  \nonumber \\
 = & \int DJ \, e^{\,\, \sum\limits^{\infty}_{r=3} \int dx_1 \cdots dx_r \frac{ (-1)^r}{r!}  G^{(r)}_{c}(x_1, \cdots , x_r) \, \partial_{\underline{r}}   }  e^{- i  \int dx\, J(x) \phi(x)  } \nonumber \\
 &e^{ i \int dx_1\, J(x_1)G^{(1)}_{c}(x_1) - \frac{1}{2}  \int dx_1\, dx_2\, J(x_1) G^{(2)}_{c}(x_1, x_2) J(x_2)  },
\end{align}
where  $\partial_{\underline{r}} = \frac{\delta}{\delta \phi(x_{1}) } \cdots \frac{\delta}{\delta \phi(x_{r}) }$. Next, we evaluate the integral associated with the Gaussian process,
\begin{align}\int DJ e^{- i  \int dx_1\, J(x_1) \phi(x_1)  + i \int dx_1\, J(x_1)G^{(1)}_{c}(x_1) - \frac{1}{2}  \int dx_1\, dx_2\, J(x_1) G^{(2)}_{c}(x_1, x_2) J(x_2)  },
\end{align}
using a change of variables $J'(x) \to J(x) + i \int dx'\, G^{(2)}_{c}(x, x')^{-1}[\phi(x') -G^{(1)}_{c}(x')]$ that keeps the measure of the source $DJ \to DJ'$ invariant. We obtain 
\begin{align} \label{eqn:EdgeworthSGPint}
& - S_G = - i \int dx \, J(x) [\phi(x) - G^{(1)}_{c}(x) ] - \frac{1}{2}\int\, dx_1 dx_2 \, J(x_1)\, G^{(2)}_{c}(x_1, x_2)\, J(x_2) \nonumber \\
= & \, - \frac{1}{2}\int dx_1\, dx_2\, \big[J(x_1) + i \int dx'\, G^{(2)}_{c}(x_1, x')^{-1}[\phi(x') - G^{(1)}_{c}(x')] \big]G^{(2)}_{c}(x_1, x_2)\big[J(x_2)  \nonumber \\ 
& + i \int\, dx''\,G^{(2)}_{c}(x_2, x'')^{-1}[\phi (x'') - G^{(1)}_{c}(x'') ] \big] - \frac{1}{2}\int\, dx''\, dx'\,dx_1\,dx_2 \,[ \phi(x'') - G^{(1)}_{c}(x'')]\,  \nonumber \\
& G^{(2)}_{c}(x'',x_2)^{-1}\, G^{(2)}_{c}(x_2, x_1) G^{(2)}_{c}(x_1, x')^{-1}\, [ \phi(x') - G^{(1)}_{c}(x')] \nonumber \\
=& \,- \frac{1}{2}\int dx_1\, dx_2\, J'(x_1)\, G^{(2)}_{c}(x_1, x_2)\, J' (x_2) - \frac{1}{2}\int\, dx\, dx' \,[ \phi(x) - G^{(1)}_{c}(x)]\, G^{(2)}_{c}(x, x')^{-1}\, \nonumber \\
&[ \phi(x') - G^{(1)}_{c}(x')]
\end{align}
An integration over $J'$ results in the distribution $$e^{ - \frac{1}{2}\int\, dx\, dx' \,[ \phi(x) - G^{(1)}_{c}(x)]\, G^{(2)}_{c}(x, x')^{-1}\, [ \phi(x') - G^{(1)}_{c}(x')]},$$ such that
\begin{align} \label{eqn:edgeappendixE}
P[\phi] &= e^{\,\,\,  \sum\limits_{r=3}^{\infty} \int dx_1 \cdots dx_r\, \frac{(-1)^r}{r!}   G^{(r)}_{c}(x_1,\cdots, x_{r} ) \,   \partial_{\underline{r}} } e^{ - \frac{1}{2}\int\, dx\, dx' \,[ \phi(x) - G^{(1)}_{c}(x)]\, G^{(2)}_{c}(x, x')^{-1}\, [ \phi(x') - G^{(1)}_{c}(x')]}. 
\end{align} 
We obtain perturbative corrections around the Gaussian field density by expanding the first exponential term in \eqref{eqn:edgeappendixE} as a series; contributions from higher order cumulants become increasingly less significant in most cases. 

\paragraph{$4$-pt Function at Finite $N$, Non-i.i.d. Parameters \label{app:secmaintextderivation}}
Next, we evaluate the $4$-pt function of this NNFT with the following cumulant generating functional
\be
    W_{\phi}[J] = \log \Bigg[ e^{W_{\phi,\vec\alpha=0}[J] } + \sum_{r=1}^{\infty} \sum_{s_1, \cdots, s_r = 1}^{q} \frac{ \alpha_{s_1} \cdots \alpha_{s_r} }{r!}\prod_{i=1}^{N}\mathbb{E}_{P_i(h_i)} \Big[e^{ \frac{1}{\sqrt{N}}\int d^dx \, h_i(x)J(x)  } \cdot \mathcal{P}_{r, \{s_1, \cdots , s_r \} } \big|_{\vec{\alpha} = 0} \Big] \Bigg].
\ee
For appropriately small $\vec{\alpha}$, the ratio of the second term in the logarithm to the first  is small, and one can Taylor expand $\log(1+ x) \approx x$ to obtain,
\begin{align}
    W_{\phi}[J] = W_{\phi,\vec\alpha=0}[J]  + \sum_{s=1}^{q} \frac{\alpha_s}{e^{W_{\phi,\vec\alpha=0}[J] }}\prod_{i=1}^{N}\mathbb{E}_{P_i(h_i)} \Big[e^{ \frac{1}{\sqrt{N}}\int d^dx \, h_i(x)J(x)  } \cdot \mathcal{P}_{1, s } \big|_{\vec{\alpha} = 0} \Big] .
\end{align}
The $4$-pt function is obtained as $G_{{c}}^{(4)} (x_1, \cdots , x_{4}) = \frac{\partial^4 W_{\phi}[J]}{\partial J(x_1) \cdots \partial J(x_4)}\big|_{J=0}$. We abbreviate
\begin{align}
    M = \sum_{s=1}^{q} \frac{\alpha_s}{e^{W_{\phi,\vec\alpha=0}[J] }}\prod_{i=1}^{N}\mathbb{E}_{P_i(h_i)} \Big[e^{ \frac{1}{\sqrt{N}}\int d^dx \, h_i(x)J(x)  } \cdot \mathcal{P}_{1, s } \big|_{\vec{\alpha} = 0} \Big],
\end{align}
then, $G_{{c}}^{(4)} (x_1, \cdots , x_{4}) = \frac{\partial^4 W_{\phi,\vec\alpha=0}[J]}{\partial J(x_1) \cdots \partial J(x_4)}\Big|_{J=0} + \frac{\partial^4 M}{\partial J(x_1) \cdots \partial J(x_4)}\Big|_{J=0}$.

Next, we evaluate the fourth $J$-derivative of $M$ and turn the source $J$ off,
\begin{equation} 
\begin{split}
    &\frac{\partial^4 M}{\partial J_1\cdots \partial J_4} \Big|_{J=0} \nonumber \\
    &= \sum_{s=1}^{q} \frac{\alpha_s}{e^{W_{\phi,\vec\alpha=0}[J] }} \Bigg( \prod_{i=1}^{N}\mathbb{E}_{P_i(h_i)} \Big[\int d^dx_1 \cdots d^dx_4 \frac{h_i(x_1) \cdots h_i(x_4)}{N^2}e^{ \frac{1}{\sqrt{N}}\int d^dx \, h_i(x)J(x)  } \mathcal{P}_{1,s } \big|_{\vec{\alpha} = 0} \Big] \nonumber \\
     &+ \sum\limits_{\mathcal{P}(abce)}\Bigg[ \Big(  \frac{\partial W_{\phi,\vec\alpha=0}[J]}{\partial J_{a}} \frac{\partial W_{\phi,\vec\alpha=0}[J]}{\partial J_{b}} - \frac{\partial^2 W_{\phi,\vec\alpha=0}[J]}{ \partial J_{a} \partial J_{b}} \Big) \prod_{i=1}^{N}\mathbb{E}_{P_i(h_i)} \Big[\int d^dx_{c}d^dx_{e} \frac{h_i(x_{c})h_i(x_{e})}{N} \nonumber \\
    & \cdot e^{ \frac{1}{\sqrt{N}}\int d^dx \, h_i(x)J(x)  } \cdot \mathcal{P}_{1,s } \big|_{\vec{\alpha} = 0} \Big] - \Big(  \frac{\partial W_{\phi,\vec\alpha=0}[J]}{\partial J_{a}} \frac{\partial W_{\phi,\vec\alpha=0}[J]}{\partial J_{b}}\frac{\partial W_{\phi,\vec\alpha=0}[J]}{\partial J_{c}} - \frac{\partial^2 W_{\phi,\vec\alpha=0}[J]}{ \partial J_{a} \partial J_{b}}  \nonumber \\
    & \cdot \frac{\partial W_{\phi,\vec\alpha=0}[J]}{\partial J_{c}} + \frac{\partial^3 W_{\phi,\vec\alpha=0}[J]}{\partial J_{a} \partial J_b \partial J_c} \Big) \prod_{i=1}^{N}\mathbb{E}_{P_i(h_i)} \Big[\int d^dx_e \frac{h_i(x_e)}{\sqrt{N}}e^{ \frac{1}{\sqrt{N}}\int d^dx \, h_i(x)J(x)  }  \mathcal{P}_{1,s } \big|_{\vec{\alpha} = 0} \Big] \nonumber \\
    & + \Big( \frac{\partial W_{\phi,\vec\alpha=0}[J]}{\partial J_{1}} \frac{\partial W_{\phi,\vec\alpha=0}[J]}{\partial J_{2}}\frac{\partial W_{\phi,\vec\alpha=0}[J]}{\partial J_{3}} \frac{\partial W_{\phi,\vec\alpha=0}[J]}{\partial J_{4}} + \frac{\partial^2 W_{\phi,\vec\alpha=0}[J]}{ \partial J_{a} \partial J_{b}} \frac{\partial^2 W_{\phi,\vec\alpha=0}[J]}{ \partial J_{c} \partial J_{e}} \nonumber \\
    &  - \frac{\partial W_{\phi,\vec\alpha=0}[J]}{\partial J_{a}}\frac{\partial W_{\phi,\vec\alpha=0}[J]}{\partial J_{b}} \frac{\partial^2 W_{\phi,\vec\alpha=0}[J]}{ \partial J_{c} \partial J_{e}} + \frac{\partial^3 W_{\phi,\vec\alpha=0}[J]}{ \partial J_{a} \partial J_{b} \partial J_c} \frac{\partial W_{\phi,\vec\alpha=0}[J]}{\partial J_{e}} - \frac{\partial^4 W_{\phi,\vec\alpha=0}[J]}{ \partial J_{1} \partial J_{2} \partial J_3 \partial J_4} \Big)  \nonumber \\
    &\prod_{i=1}^{N}\mathbb{E}_{P_i(h_i)} \Big[e^{ \frac{1}{\sqrt{N}}\int d^dx \, h_i(x)J(x)} \mathcal{P}_{1,s } \big|_{\vec{\alpha} = 0} \Big] -  \prod_{i=1}^{N}\mathbb{E}_{P_i(h_i)} \Big[\int d^dx_b d^dx_c d^dx_e \frac{h_i(x_b) h_i(x_c)h_i(x_e)}{N^{3/2}} \nonumber \\
    & e^{ \frac{1}{\sqrt{N}}\int d^dx \, h_i(x)J(x)  } \mathcal{P}_{1,s } \big|_{\vec{\alpha} = 0} \Big] \frac{\partial W_{\phi,\vec\alpha=0}[J]}{\partial J_{a}}  \Bigg] \Bigg) \Bigg|_{J=0} := \vec{\alpha} \cdot \Delta G^{(4)}_{c} (x_1, \cdots , x_{4}) ,
\end{split}
\end{equation}
where we use the abbreviation $J(x_i):= J_i$. In the mean-free case, 
\begin{align}
    \frac{\partial W_{\phi,\vec\alpha=0}[J]}{\partial J_{a}}\Big|_{J=0} &= \frac{\partial^3 W_{\phi,\vec\alpha=0}[J]}{\partial J_{a} \partial J_{b} \partial J_c}\Big|_{J=0} = 0, \\
    \frac{\partial^4 W_{\phi,\vec\alpha=0}[J]}{\partial J_{1} \partial J_{2} \partial J_3 \partial J_4}\Big|_{J=0} &= G_{{c}}^{(4), \text{i.i.d.}} (x_1, \cdots , x_{4}), ~~
    \frac{\partial^2 W_{\phi,\vec\alpha=0}[J]}{\partial J_{a} \partial J_{b} } \Big|_{J=0} = G_{{c}}^{(2), \text{i.i.d.}} (x_a , x_b).
\end{align} 
Thus, the $4$-pt function is
\begin{align}
    &G_{{c}}^{(4)} (x_1, \cdots , x_{4}) = G_{{c}}^{(4), \text{i.i.d.}} (x_1, \cdots , x_{4}) \nonumber \\
    &\qquad + \sum_{s=1}^{q} \frac{\alpha_s}{e^{W_{\phi,\vec\alpha=0}[J=0] }} \Bigg( \prod_{i=1}^{N}\mathbb{E}_{P_i(h_i)} \Big[\int d^dx_1 \cdots d^dx_4 \frac{h_i(x_1) \cdots h_i(x_4)}{N^2} \mathcal{P}_{1,s } \big|_{\vec{\alpha} = 0} \Big] \nonumber \\
    &\qquad + \sum\limits_{\mathcal{P}(abce)}\Big[  - G_{{c}}^{(2), \text{i.i.d.}} (x_a , x_b) \prod_{i=1}^{N}\mathbb{E}_{P_i(h_i)} \Big[\int d^dx_{c}d^dx_{e} \frac{h_i(x_{c})h_i(x_{e})}{N}   \mathcal{P}_{1,s } \big|_{\vec{\alpha} = 0} \Big]  \nonumber \\
    & \qquad + \Big(  G_{{c}}^{(2), \text{i.i.d.}} (x_a , x_b)G_{{c}}^{(2), \text{i.i.d.}} (x_c , x_e)  - G_{{c}}^{(4), \text{i.i.d.}} (x_1, \cdots , x_{4}) \Big) \prod_{i=1}^{N}\mathbb{E}_{P_i(h_i)} \Big[ \mathcal{P}_{1,s } \big|_{\vec{\alpha} = 0} \Big] \Big]\Bigg), \nonumber \\
    & \qquad = G_{{c}}^{(4), \text{i.i.d.}} (x_1, \cdots , x_{4}) + \vec{\alpha} \cdot \Delta G^{(4)}_{c} (x_1, \cdots , x_{4}) + O(\vec{\alpha}^2). \label{app:maintextderivation}
\end{align}
at leading order.

\section{Fourier Transformation Trick for $G^{(2)}_c(x,y)^{-1}$ \label{app:G2inverseFT}}
Let us evaluate the  expression
\begin{align}
    & \int dy_1 \cdots dy_{n} \, G_{{c}}^{(n)} (y_1, \cdots , y_{n}) \, G_{{c}}^{(2)}(y_1, x_1)^{-1}\cdots G_{{c}}^{(2)}(y_n, x_n)^{-1} ,
\end{align}
when $G_{{c}}^{(2)}(y_i, x_i)^{-1}$ involves differential operators. The integrals over  $y_i$ cannot be directly evaluated as the eigenvalues of each $G_{{c}}^{(2)}(y_i, x_i)^{-1}$ are unknown. To avoid this problem, we substitute the operators and cumulant with their Fourier transformations,
\begin{align}
    &\int d^dy_1\cdots d^dy_n \, d^dp_1\cdots d^dp_n \, d^dq_1\cdots d^dq_n \, d^dr_1\cdots d^dr_n \, \tilde{G}_{{c}}^{(n)} (p_1, \cdots , p_{n}) \, \tilde{G}_{{c}}^{(2)}(q_1, r_1)^{-1} \nonumber \\
    &\quad \cdots \tilde{G}_{{c}}^{(2)}(q_n, r_n)^{-1} \, e^{i y_1(p_1 + q_1) + ir_1 x_1  \cdots + i y_1(p_n + q_n) + ir_n x_n}  \nonumber \\
    &= \int d^dp_1 \cdots d^dp_n \, d^dr_1 \cdots d^dr_n \, \tilde{G}_{{c}}^{(n)} (p_1, \cdots , p_{n}) \, \tilde{G}_{{c}}^{(2)}(-p _1, r_1)^{-1} \cdots \tilde{G}_{{c}}^{(2)}(-p_n, r_n)^{-1} e^{i \sum\limits_{j=1}^{n} r_jx_j }. \label{app:useforReLU} 
\end{align}
Here $\tilde{f}$ is the Fourier transformation of $f$, and we obtained the second line by evaluating $y_i$ integrals to get $\delta^d(p_i + q_i)$, then integrating $q_i$ variables.

When $G_{{c}}^{(2)}$ is translation invariant, we have $G_{{c}}^{(2)}(y_i, x_i)^{-1} \propto \delta^d(y_i - x_i)$, leading to further simplification of the above expression as,
\begin{align}
    \int d^dp_1 \cdots d^dp_n  \, \tilde{G}_{{c}}^{(n)} (p_1, \cdots , p_{n}) \, \tilde{G}_{{c}}^{(2)}(-p_1)^{-1} \cdots \tilde{G}_{{c}}^{(2)}(-p_n)^{-1} e^{ -i p_1x_1  \cdots  - ip_nx_n }. \label{app:useforCosGauss}
\end{align}
We exemplify this expression for Cos-net and Gauss-net architectures.

\section{Gaussian Processes: Locality and Translation Invariance} \label{app:localityTinv}

Any Gaussian process (GP) can be described as a function space distribution given by action $S$,
\begin{equation}
    S = \int dx \, dy \, f(x) \, G^{(2)}_{c}(x, y)^{-1} \, f(y),
\end{equation}
where $G^{(2)}_{c}(x,y)^{-1}$ is the \textit{precision function}, related to the GP kernel by the inversion formula 
\begin{equation} \label{eq:inversion}
    \int dy \, G^{(2)}_{c}(x, y)^{-1} \, K(y, z) = \delta(x - z).
\end{equation}
A \textit{local} GP can be defined as a family of functions with a completely diagonalizable precision operator, resulting in the action
\begin{equation}
    S = \int dx \, f(x) \, G^{(2)}_{c}(x)^{-1} \, f(x),
\end{equation}
with the inversion relation simplified into 
\begin{equation} \label{eq:localinversion}
    G^{(2)}_{c}(x)^{-1} \, K(x, z) = \delta(x - z).
\end{equation}
This can be seen by considering $G^{(2)}_{c}(x, y)^{-1} = \delta(x - y) \Sigma(x)$ and performing the integral over $y$ in Eqn. \eqref{eq:inversion}. A Gaussian process can always be written in a local basis, as we will show below.

\vspace{1cm}
\noindent {\bf Gaussian Process  Action in the local basis.}
Any Gaussian Process $f(x)$, when evaluated at a discrete set of inputs $\{x_{i}\}_{i}$, forms a multivariate Gaussian distribution. The covariance matrix of a multivariate Gaussian is a real symmetric matrix, and thus can be diagonalized. We can use this diagonalization procedure on the Gaussian process distribution itself, thereby rewriting it with a kernel proportional to a Dirac delta function,
\begin{align} \label{eq:diag1}
    S &= - \frac{1}{2} \int d^{d}x_{i} \, d^{d}x_{l} \, f(x_{i}) \, G^{(2)}_{c}(x_{i}, x_{l})^{-1} \, f(x_{l}),  \nonumber \\
    &= -\frac{1}{2}\int d^d x_i d^d x_j d^d x_k d^d x_l f(x_i)V(x_i,x_j)D(x_j, x_k)V^{-1}(x_k, x_l)f(x_l), \nonumber \\
    &= - \frac{1}{2}\int d^d x_k \left[\int d^d x_i \, V(x_i, x_k) f(x_k)\right] \Sigma(x_k) \left[\int d^d x_l \, V^{-1}(x_k, x_l) f(x_l)\right], \nonumber \\
    &= -\frac{1}{2} \int d^dx \, \phi^T(x) \Sigma(x) \phi(x),
\end{align}
where $\phi(x) := \int d^d y \,f(y) V(y,x)$ and last step of \eqref{eq:diag1} is obtained by $x_k \xrightarrow[]{} x$. $D(x,y)$ is defined as $D(x_i,x_l) = \delta(x_i-x_l)\Sigma(x_i) = \int d^d x_j d^d x_k \, V^{-1}(x_i,x_j) G^{(2)}_{c}(x_j, x_k)^{-1} V(x_k, x_l)$.

\end{appendices}

\bibliography{refs}

\end{document}